\documentclass{article}
\usepackage{authblk}
\usepackage{geometry}
\usepackage{graphicx} 
\usepackage{amsmath}
\DeclareMathOperator{\Tr}{Tr}
\DeclareMathOperator{\inv}{inv}
\DeclareMathOperator{\sgn}{sgn}
\usepackage{amssymb}
\usepackage{multirow}
\usepackage{bm}
\usepackage{upgreek}
\usepackage[edges]{forest}
\usepackage{parskip}
\usepackage{orcidlink}
\usepackage{xcolor}
\hypersetup{
    colorlinks,
    linkcolor={red!50!black},
    citecolor={blue!50!black},
    urlcolor={blue!80!black}
}

\sffamily

\title{The link between Microstructural Heterogeneity, Diffusivity, and Hydrogen Embrittlement\\\rule[0.2cm]{15cm}{0.4pt}}
\author[]{Daniel J. Long$^*$\orcidlink{0009-0004-2068-7871}}
\author[]{Edmund Tarleton\orcidlink{0000-0001-6725-9373}}
\author[]{Alan C.F. Cocks\orcidlink{0000-0002-6245-3406}}
\author[]{Felix Hofmann$^{\dag}$\orcidlink{0000-0001-6111-339X}}
\affil[]{Department of Engineering Science, University of Oxford, Parks Road, OX1 3PJ Oxford, UK}
\begin{document}
\date{}
\maketitle
\centerline{Correspondence: $^*$\href{mailto:daniel.long@eng.ox.ac.uk}{daniel.long@eng.ox.ac.uk}, $^{\dag}$\href{mailto:felix.hofmann@eng.ox.ac.uk}{felix.hofmann@eng.ox.ac.uk}}

\section*{Abstract}
Green hydrogen is likely to play a major role in decarbonising the aviation industry. It is crucial to understand the effects of microstructure on hydrogen redistribution, which may be implicated in the embrittlement of candidate fuel system metals. We have developed a stochastic multiscale finite element modelling framework that integrates micromechanical and hydrogen transport models, such that the dominant microstructural effects can be efficiently accounted for at millimetre length scales. Our results show that microstructure has a significant effect on hydrogen localisation in elastically anisotropic materials, which exhibit an interesting interplay between microstructure and millimetre-scale hydrogen redistribution at various loading rates. Considering 316L stainless steel and nickel, a direct comparison of model predictions against experimental hydrogen embrittlement data reveals that the reported sensitivity to loading rate is strongly linked with rate-dependent grain scale diffusion. These findings highlight the need to incorporate microstructural characteristics in the design of hydrogen resistant materials.

\section{Introduction}
As the UK government strives to achieve Net Zero by 2050, many of the largest industries are exploring pathways to integrate renewable and environmentally friendly energy sources. For the aerospace sector, hydrogen may become a viable alternative propulsion fuel. The challenges of this, however, are the known damaging effects of hydrogen in metallic materials \cite{COTTERILL1961}. This is not to mention the unprecedented complexity of a hydrogen fuel system which would encompass vast temperature and pressure ranges (from high pressure cryogenic storage to high temperature combustion \cite{PALIES2022}). Whilst many models \cite{CUI2017,HASSAN2019} have been developed to capture hydrogen embrittlement (HE) in structural metallic materials, little has been achieved towards explaining the large degree of variation reported between different alloys \cite{DWIVEDI2018}. Understanding this will be vital for effective material selection and design. Using a multiscale modelling methodology, this paper investigates the connection between varying degrees of HE and relevant microstructural properties. 

Across the theoretical and experimental literature, the most widely reported mechanisms for HE (in non-hydride forming materials) are hydrogen enhanced localised plasticity (HELP) \cite{KATZAROV2017I,DECONINCK2024} and hydrogen enhanced decohesion (HEDE) \cite{KATZAROV2017II,TAKAHASHI2016}. The mechanistic basis for HELP is that dislocations sequester and trap mobile hydrogen, which enhances dislocation mobility, resulting in the accelerated accumulation of dislocations, i.e., localised plasticity and damage \cite{MAO2022}. Existing experimental evidence \cite{ROBERTSON1999}, which is supported by discrete dislocation analyses \cite{YU2019}, suggests that hydrogen reduces the interaction energies between dislocations, leading to diminished hardening rates and the early onset of plasticity. The HEDE mechanism suggests that the interfacial cohesive strengths of grain boundaries (GBs) and other microstructural hydrogen trapping sites are diminished by the presence of hydrogen, resulting in brittle interfacial fracture (as opposed to ductile void growth and coalescence), for which there is ample experimental evidence \cite{ABOURA2021,OKAYASU2020,KIMURA1988}. Atomistic modelling by Katzarov et al. \cite{KATZAROV2017II} showed that during decohesion of (111) planes in stressed and hydrogenated $\alpha$-Fe, plane separation promotes hydrogen ingress and trapping (due to expansion of interstitial sites), which reduces the interface binding energy and hence the interfacial cohesive strength by up to a third, even for very low bulk hydrogen concentrations. While HELP and HEDE mechanisms are often attributed to the embrittlement of materials independently of one another, it is conceivable that they act reciprocally, viz., HELP promotes HEDE. Robertson et al. \cite{ROBERTSON2015} reported that HELP accelerates the rate at which glissile dislocations pile up at GBs, while also reducing pile-up dislocation spacing (due to hydrogen-affected dislocation interaction energies). The consequences of increased dislocation mobility and density are that (i) there is an increased probability of slip along GBs, which can weaken the cohesive strength, (ii) transport of hydrogen to the GBs is increased via high density dislocations networks, and (iii) GB disorder will increase due to the increased likelihood for slip transfer between grains where the local stress field is enhanced by dense dislocation pile-ups \cite{ROBERTSON2015,QI2007}. The main consequence of (iii) is that the capacity for hydrogen at GBs will be increased. These HE mechanisms are strongly influenced by the local hydrogen content, and hence, factors which affect the redistribution of hydrogen at microstructure length scales need to be carefully considered. Hassan et al. \cite{HASSAN2019} used a micromechanical crystal plasticity finite element (CPFE) modelling framework to study the transient redistribution of hydrogen in pre-hydrogenated non-textured polycrystals under external and plasticity-driven residual internal stresses after deformation. The diffusion model accounted for the evolution of hydrogen traps (dislocations) driven by plastic strain evolution, which were shown to substantially reduce the overall apparent diffusivity through polycrystalline 316L stainless steel. Moreover, the application of 20\% biaxial strain to the equiaxed polycrystal was predicted to drive H concentrations up by up to a factor of two (relative to the initial value), while residual stresses due to plasticity in the unloaded state yielded similar results. Tondro et al. \cite{TONDRO2023} adopted a similar framework to study the hydrostatic stress-driven redistribution of hydrogen in zirconium polycrystals; results showed that the distribution of loads between hard and soft grains significantly influences how hydrogen is distributed. Specifically, load shedding \cite{VENKATRAMANI2007} means that higher hydrostatic stresses are generated in hard grains, driving an increased hydrogen-influx, whilst soft grains exhibit more plasticity and greater dislocation content which increase trapped hydrogen. In general, GB stress concentration features accumulated high levels of hydrogen, and triple junctions were found to promote particularly high concentrations, likely to foster HEDE. 

In modelling the direct mechanical effects of hydrogen, numerous approaches from atomistic to millimetre length scales have been utilised. Ab initio models \cite{KATZAROV2017I,KATZAROV2017II} were previously discussed in the context of HE; these methods have generated important mechanistic insights into HELP and HEDE, which are crucial for deterministic model development at larger length scales. Similarly, dislocation-level models have been developed to study these respective embrittlement mechanisms. Using a discrete dislocation plasticity (DDP) framework, Irani et al. \cite{IRANI2017} showed that the tensile hydrostatic stress field generated by an individual dislocation (trap) attracts hydrogen; Yu et al. \cite{YU2019} later demonstrated via DDP modelling that hydrogen can accelerate dislocation generation and slip planarity. While both of these findings provide helpful reinforcing evidence of HELP, due to the pragmatic limitations of high-fidelity modelling, DDP cannot yet predict material behaviour at component length scales. At the grain scale, CPFE modelling has been most widely used to study microstructure-driven hydrogen redistribution and embrittlement \cite{HASSAN2019,ILIN2014,ABDOLVAND2019,HUSSEIN2021}. Ilin et al. \cite{ILIN2014} were among the first to illustrate the effects of microstructure heterogeneity on the localisation of hydrogen in a case study of 316L stainless steel under uniaxial tensile loads; strain-rate was shown to impact the magnitude of redistribution significantly due to the time dependent nature of hydrogen diffusion. Hussein et al. \cite{HUSSEIN2021} later presented a thorough comparative study of the micromechanical redistribution of hydrogen in high and low-strength ferritic steels exposed to various hydrogen pressure conditions. A statistical analysis after uniaxial tensile loading of a 2D (plane strain) polycrystal with random texture revealed that diffusible hydrogen followed a Gaussian distribution, while trapped hydrogen was better characterised by a log-normal distribution, owing to the development of highly localised plasticity (and hence dislocation traps), driven by asymptotic stress fields. However, as the loading rate considered was very low compared to the time required for diffusion, the resultant hydrogen distributions were effectively at steady state. Moreover, the factors affecting the spread and shape of the distributions were not elucidated. Therein lies a key literature gap which needs to be addressed to (i) uncover the role of microstructure heterogeneity and local hydrogen transport in affecting HE, and (ii) develop microstructure-sensitive models at millimetre length scales. This is particularly relevant for industrial applications, where microstructural effects are highly important, but also where component length scales exceed the practical limitations of e.g., CPFE modelling. 

Based on known marked differences in hydrogen diffusivity between metals with body-centred cubic (BCC) and face-centred cubic (FCC) crystallography (orders of magnitude lower) \cite{MARTIN2020}, we hypothesise that the reportedly higher susceptibility of e.g., ferritic steels to HE is linked to higher rates of diffusion and localisation of hydrogen at the microstructural level than in e.g., austenitic steels \cite{PERNG1987}. To address (i) and (ii), this paper presents the development of a new multiscale modelling methodology which explicitly accounts for the local elastic interactions that promote hydrogen redistribution in a rate-dependent manner. By coupling new insights from CPFE modelling with a continuum level mechanical and hydrogen diffusion modelling framework \cite{ELMUKASHFI2020} for a direct comparison against published experimental data, the following question is addressed: To what extent does microstructure heterogeneity promote the redistribution of hydrogen and associated HE in metals? 

\section{Methodology}
\subsection{Micromechanical model} \label{MicroModel}

To study grain scale hydrogen redistribution, a dislocation-based CPFE modelling framework \cite{DEMIR2025,HARDIE2023} is used. Only the mechanical model is presented here, as a description of the diffusion model is given in subsection \ref{ChemPot}. 316L stainless steel is used here as a case study as it is well characterised in the literature and is a candidate material for hydrogen fuel system and storage applications \cite{LIU2023}. The CPFE constitutive laws are formulated for large deformations, and adopt Lee's multiplicative decomposition of the deformation gradient \cite{LEE1969} as follows, 

\begin{equation}
\bm{F}=\bm{F}^{\mathrm{e}}\bm{F}^{\mathrm{p}}\label{F}
\end{equation}

where $\bm{F}$, $\bm{F}^{\mathrm{e}}$, and $\bm{F}^{\mathrm{p}}$ are the total, elastic, and plastic deformation gradients, respectively. The elastic deformation gradient is obtained from the total and plastic contributions ($\bm{F}^{\mathrm{e}}=\bm{F}{\bm{F}^{\mathrm{p}}}^{-1}$), and is used to determine the stress tensor via Hooke's law. Within an implicit integration scheme, the elastic strain increment is obtained from $\Delta\bm{\epsilon}^{\mathrm{e}}=\mathrm{sym}(\bm{L}^{\mathrm{e}})\Delta t$, where $\bm{L}^{\mathrm{e}}$ is the elastic velocity gradient and $\Delta t$ is the time increment. Assuming room temperature plastic deformation in 316L is controlled by dislocation glide, the following mechanistic slip rule \cite{DUNNE2007} is applied to each slip system, $i\in[1,M]$ ($M=12$ for FCC materials). Equation (\ref{gamma_i}) derives from Gibbs' statistical mechanics description of the thermally-activated escape of pinned dislocations \cite{GIBBS1964}.

\begin{equation}
\dot{\gamma}^i=\rho_{\mathrm{m}}v{b^i}^2\exp\left(\frac{-\Delta F}{k_{\mathrm{B}}T}\right) \sinh\left(\frac{|\tau^i|-\tau_{\mathrm{c}}^i)\Delta V}{k_{\mathrm{B}}T}\right)\sgn{(\tau^i)}\label{gamma_i}
\end{equation}

\begin{figure}[htb]
  \centering
  \includegraphics[width=0.95\linewidth]{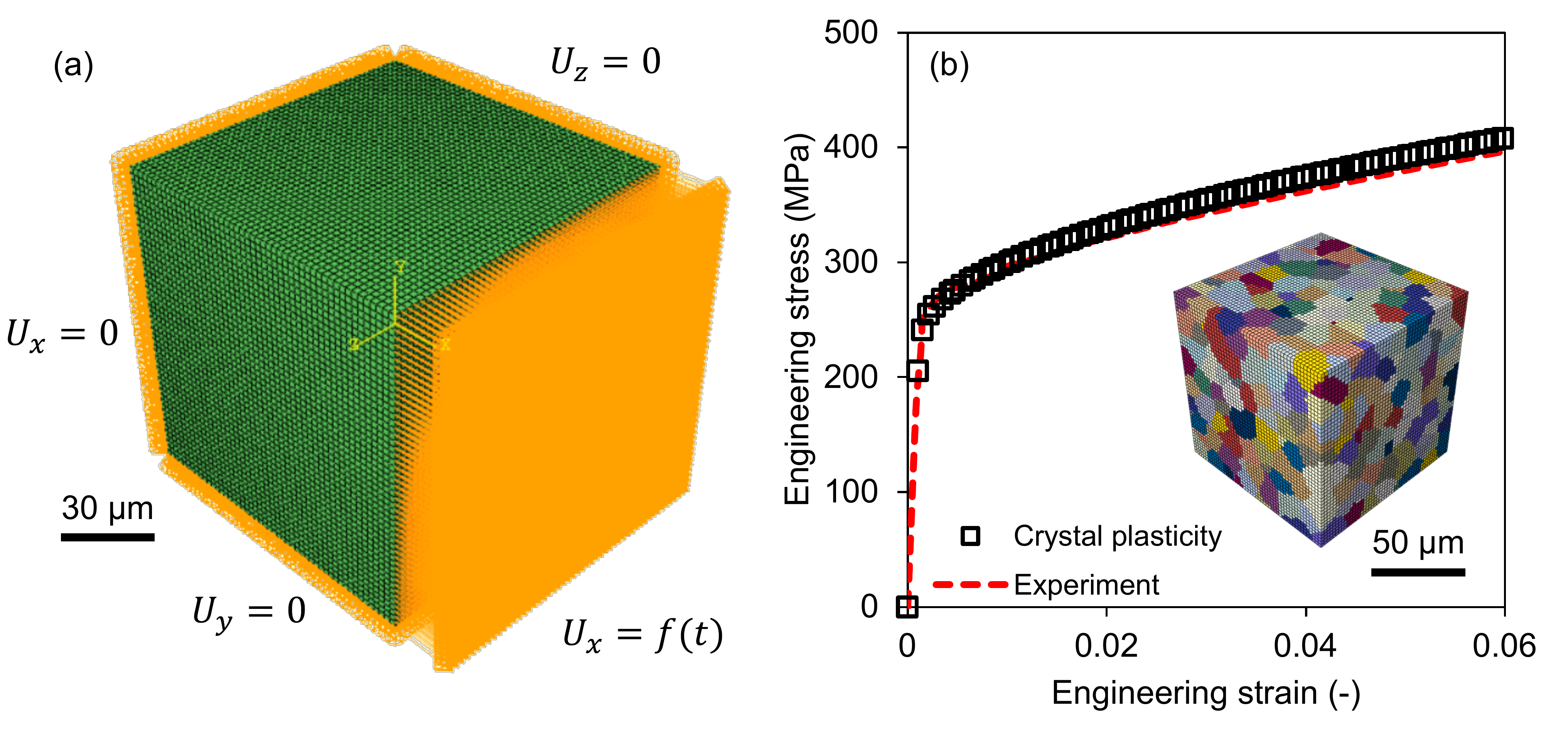}
  \caption{Overview of the CPFE model. (a) Mesh and boundary conditions imposed on a 100 $\times$ 100 $\times$ 100 $\upmu$m$^3$ representative polycrystal model. (b) Comparison of bulk mechanical response of polycrystal model (using 316L properties) with 316L experimental data \cite{KANG2010}. The 10 $\upmu$m and 20 $\upmu$m grain diameter models yielded identical mechanical responses. A subfigure of the 291 grain polycrystal model is shown, in which each colour represents a grain with unique crystallographic orientation.}
  \label{modelBCs}
\end{figure}

$\dot{\gamma}^i$ is the plastic shear strain-rate, $\rho_{\mathrm{m}}$ the density of mobile dislocations (assumed constant in the absence of dislocation generation and pinning data), $v$ represents the frequency of attempts for dislocations to overcome thermal barriers, $b^i$ the slip system Burgers vector magnitude, $\Delta F$ the activation energy, $\tau^i$ and $\tau_{\mathrm{c}}^i$ the applied and critical resolved shear stresses for slip system $i$, $\Delta V$ the activation volume, and $k_{\mathrm{B}}$ and $T$ represent Boltzmann constant and temperature, respectively. The full plastic strain tensor may then be obtained at each Gauss point by numerical integration of the plastic velocity gradient, $\bm{L}^{\mathrm{p}}$, which is given by Schmid's law as follows, 

\begin{equation}
\bm{L}^{\mathrm{p}}=\sum_{i=1}^{M}(\dot{\gamma}^i\bm{s}^i\otimes\bm{n}^i) \label{Lp}
\end{equation}

where $\bm{s}^i$ and $\bm{n}^i$ represent slip direction and slip plane unit normal vectors, respectively. The plastic deformation gradient is related to the plastic velocity gradient by $\bm{L}^{\mathrm{p}}=\dot{\bm{F}}^{\mathrm{p}}{\bm{F}^{\mathrm{p}}}^{-1}$. To capture plastic strain hardening in 316L, the classical Taylor model is applied to each slip system,

\begin{equation}
\tau_{\mathrm{c}}^i=\tau_{\mathrm{c,0}}^i+G^ib^i\sqrt{\rho_{\mathrm{SSD}}} \label{tau_c}
\end{equation}

where $\tau_{\mathrm{c,0}}^i$ is the critical resolved shear stress in the absence of dislocation obstacles, $G^i$ the shear modulus in the dislocation glide plane, and $\rho_{\mathrm{SSD}}$ the total density of statistically stored dislocations (SSDs). As an objective of this work is to develop analytical models that capture microstructure length scale effects on the redistribution of hydrogen, the secondary effect of geometrically necessary dislocation (GND) hardening is not considered. Nevertheless, the linear SSD evolution law given by Equation (\ref{rho_ssd}) captures the experimental uniaxial tensile behaviour of polycrystalline 316L \cite{KANG2010} well, as shown in Figure \ref{modelBCs}.

\begin{equation}
\rho_{\mathrm{SSD}}=\rho_{\mathrm{SSD,0}}+kp^{\mathrm{local}} \label{rho_ssd}
\end{equation}

In Equation (\ref{rho_ssd}), $\rho_{\mathrm{SSD,0}}$ represents the initial density of SSDs, $k$ is a hardening constant, and $p^{\mathrm{local}}$ is cumulative plastic strain, which is given by Equation (\ref{p}) \cite{DUNNE2005}.

\begin{equation}
p^{\mathrm{local}}=\left(\frac{2}{3}{\bm{\epsilon}}^{\mathrm{p}}:{\bm{\epsilon}}^{\mathrm{p}}\right)^{\frac{1}{2}} \label{p}
\end{equation}

This constitutive model is applied to two polycrystalline representative volume elements (RVEs) with equiaxed grains with diameters of 10 $\upmu$m and 20 $\upmu$m, respectively. The RVE boundary conditions are shown in Figure \ref{modelBCs} (a), with displacement-controlled loading applied over various time periods. Though Equation (\ref{gamma_i}) is formulated to capture rate-dependent plasticity, the mechanical response of 316L is virtually insensitive to strain-rate in the low strain-rate regime considered here (${\dot{\epsilon}}_x\leq10^{-3}$ s$^{-1}$) \cite{CHOUDHARY2014}. For each RVE, the element size is set to one tenth of the mean grain diameter. The resultant models, which were generated using the open source DREAM 3D software package \cite{DREAM3D}, were assigned random grain orientations (non-textured), and consist of 291 and 250 grains, corresponding to models with 10 $\upmu$m and 20 $\upmu$m grain diameters, respectively. A comparison of the simulated and experimental room temperature ($T=293$ K) mechanical response up to 6\% strain is presented in Figure \ref{modelBCs}. 

Applying cubic symmetry, the crystallographic elastic properties for 316L \cite{LEDBETTER1984} are presented in Table \ref{MicroProps}, along with the slip rule and Taylor hardening parameters from Barzdajn et el. \cite{BARZDAJN2018}.

\begin{table}[h!]
\centering
\caption{Elastic-plastic micromechanical properties for 316L.}
\begin{tabular}{ l l l }
\hline 
 Property                          & Value                & Units      \\
 \hline
 \multicolumn{3}{l}{Elastic properties \cite{LEDBETTER1984}}                                \\
 \hline
 $C_{11}$                          & $205\times10^{9}$    & Pa         \\ 
 $C_{12}$                          & $138\times10^{9}$    & Pa         \\ 
 $C_{44}$                          & $126\times10^{9}$    & Pa         \\ 
 $\nu_{12}$                        & $0.402$              & -          \\
 \hline
 \multicolumn{3}{l}{Slip rule properties \cite{BARZDAJN2018}}                              \\
 \hline
 $\tau_{\mathrm{c},0}^{i\in[1,M]}$ & $80\times10^{6}$     & Pa         \\
 $\rho_{\mathrm{m}}$               & $2.45\times10^{10}$  & m$^{-2}$   \\
 $v$                               & $10^{11}$            & s$^{-1}$   \\
 $b^{i\in[1,M]}$                   & $2.54\times10^{-10}$ & m          \\
 $\Delta F$                        & $2.6\times10^{-20}$  & J          \\
 $\Delta V$                        & $25.2$               & $b^3$      \\
  \hline
 \multicolumn{3}{l}{Taylor model properties \cite{BARZDAJN2018}}                           \\
 \hline
 $\rho_{\mathrm{SSD},0}$           & $2.45\times10^{10}$  & m$^{-2}$   \\
 $G^{i\in[1,M]}$                   & $126\times10^{9}$    & Pa         \\
 $k$                               & $7\times10^{13}$     & m$^{-2}$   \\
 \hline
\end{tabular}
\label{MicroProps}
\end{table}

\subsection{Continuum mechanical model} \label{MacroModel}

The continuum mechanical model is based on a von Mises yield criterion, assuming rate-independent isotropic elastic-plastic behaviour, and is applicable to monotonic loading conditions. Under fully elastic loading, stress is obtained via Hooke's law. The von Mises stress is given by,

\begin{equation}
\sigma_{\mathrm{VM}}=\left(\frac{3}{2}{\bm{\sigma}}':\bm{\sigma}'\right)^{\frac{1}{2}} \label{mises}
\end{equation}

where $\bm{\sigma}'$ is the deviatoric stress tensor. When the von Mises stress exceeds the material yield stress ($\sigma_{\mathrm{VM}}\geq\sigma_{\mathrm{y}}$), the Taylor model from section \ref{MicroModel} is adopted to capture hardening at the continuum level. Hence, flow stress is given by,

\begin{equation}
\sigma_{\mathrm{y}}=\sigma_{\mathrm{y}}^0+M_{\mathrm{T}}G^i(0.5b^i\sqrt{\rho_{\mathrm{SSD}}}) \label{sigma_y}
\end{equation}

where $\sigma_{\mathrm{y}}^0$ is the initial yield strength of the undeformed material (excluding the initial dislocation hardening contribution), $M_{\mathrm{T}}$ is the Taylor factor (accounting for the relationship between shear and normal stresses in non-textured FCC polycrystals, since $\sigma=M_{\mathrm{T}}\tau$ \cite{ZHANG2019}), and $\epsilon_{\mathrm{SSD}}^{\mathrm{e}}=0.5\gamma_{\mathrm{SSD}}^{\mathrm{e}}=0.5b^i\sqrt{\rho_{\mathrm{SSD}}}$ represents the cumulative elastic shear strain generated by SSDs, i.e., factoring the difference between mathematical and engineering shear strains. SSD density evolution is also governed by the same linear law as before,

\begin{equation}
\rho_{\mathrm{SSD}}=\rho_{\mathrm{SSD,0}}+kM_{\mathrm{T}}p^{\mathrm{cont}} \label{rho_ssd2}
\end{equation}

where $p^{\mathrm{cont}}$ is plastic strain at the continuum level. Note the difference between Equations (\ref{rho_ssd}) and (\ref{rho_ssd2}) is that the Taylor factor is incorporated here to maintain consistency between local and continuum strain terms, since $p^{\mathrm{cont}}=p^{\mathrm{local}}/M_{\mathrm{T}}$ \cite{ZHANG2019}. In section \ref{results}, the continuum mechanical model is applied to case studies of 316L and pure nickel. Their respective mechanical and hydrogen transport properties are also provided there.

\subsection{Hydrogen transport model} \label{ChemPot}

The hydrogen transport framework was originally presented by Elmukashfi et al. \cite{ELMUKASHFI2020} for continuum level coupled mechanical and diffusion problems. An extension is presented with the micromechanical framework (section \ref{MicroModel}) for microscale diffusion analyses. The hydrogen transport code is used in this paper for both the micromechanical and continuum mechanical models. The framework is formulated to account for hydrogen traps, including dislocations and fixed traps, e.g., grain or precipitate boundaries or interfaces. Only dislocation traps are considered in this paper. At a material point (continuum or local), the total hydrogen concentration, $C$, is therefore given by,

\begin{equation}
C=C_{\mathrm{L}}+C_{\mathrm{T}}\label{C_total}
\end{equation}

where $C_{\mathrm{L}}$ is the concentration of (diffusible) hydrogen in the metal lattice and $C_{\mathrm{T}}$ is the concentration of trapped hydrogen. Concentrations are given by the product of their respective occupancy fractions, $\theta \in [0,1]$, and saturation concentrations. Hence the lattice concentration of hydrogen is given by,

\begin{equation}
C_{\mathrm{L}}=\beta N_{\mathrm{L}} \theta_{\mathrm{L}}\label{C_L}
\end{equation}

where constants $\beta$ and $N_{\mathrm{L}}$ are the number of interstitial solvent sites per atom (dependent on crystallography) and the number of solvent atoms per unit volume, respectively. The product of $\beta$ and $N_{\mathrm{L}}$ give the saturation concentration of interstitial hydrogen. Similarly, the concentration of trapped hydrogen is given by,

\begin{equation}
C_{\mathrm{T}}=\sum_{j=1}^{n} \alpha_j N_j \theta_j\label{C_T}
\end{equation}

where $j$ represents the trap type, $\alpha_j$ denotes the number of trapping sites in a trap, and $N_j$ the density of trapping sites per unit volume. For fixed traps, $N_j=N_{\mathrm{F}}$ is assumed constant (and here is set to zero). For dislocation traps $\alpha_j=\alpha_{\mathrm{D}}=1$ and $N_j=N_{\mathrm{D}}$ is assumed proportional to the dislocation density, as shown in Equation (\ref{N_D}).

\begin{equation}
N_{\mathrm{D}}=\frac{1}{b}\rho_{\mathrm{SSD}}\label{N_D}
\end{equation}

Here, $\rho_{\mathrm{SSD}}$ represents dislocation density which can be a function of $p^{\mathrm{local}}$ or $p^{\mathrm{cont}}$. In Equation (\ref{N_D}), the Burgers vector magnitude, $b$, is taken along the FCC $<110>$ direction. 

Oriani's equilibrium condition \cite{ORIANI1970} allows for trap occupancies to be obtained directly from the lattice occupancy, $\theta_{\mathrm{L}}$,

\begin{equation}
\frac{\theta_j}{1-\theta_j}=\frac{\theta_{\mathrm{L}}}{1-\theta_{\mathrm{L}}}K_j\label{Oriani}
\end{equation}

where $K_j=\exp(-W_j/RT)$ is a constant; $W_j$ represents the binding energy for the $j^{\mathrm{th}}$ trap and $R$ is the universal gas constant. In modelling the transient diffusion of hydrogen through the lattice, the hydrogen flux vector is related to the spatial gradient of chemical potential as follows,

\begin{equation}
\bm{J}=-\frac{D_{\mathrm{L}}C_{\mathrm{L}}}{RT}\frac{\partial\mu}{\partial \bm{x}}\label{J}
\end{equation}

where $D_{\mathrm{L}}$ is lattice diffusivity and $\mu$ is chemical potential. The chemical potential derives from the rate of change of Gibbs energy with respect to lattice concentration and is given by,

\begin{equation}
\mu=\mu_0+RT\ln{\frac{C_{\mathrm{L}}}{C_{\mathrm{L}}^{\mathrm{max}}-C_{\mathrm{L}}}}+\mu_\sigma\label{mu}
\end{equation}

where $\mu_0$ is a constant and represents the chemical potential at some standard condition (and hence has no impact on the overall spatial gradient and is encoded here as 0), $C_{\mathrm{L}}^{\mathrm{max}}=\beta N_{\mathrm{L}}$ is the lattice saturation concentration, and $\mu_\sigma=\sigma_{\mathrm{H}}V_{\mathrm{L}}$ characterises the effect of hydrostatic stress on chemical potential. $\sigma_{\mathrm{H}}=\frac{1}{3}\Tr\boldsymbol{\sigma}$ is the hydrostatic stress and $V_{\mathrm{L}}=N_{\mathrm{A}}/N_{\mathrm{L}}$ is the molar volume of the solvent lattice ($N_{\mathrm{A}}$ is Avogadro's constant).

For each model considered here, an initial chemical potential is defined homogeneously throughout, giving a total hydrogen concentration which remains unchanged in the absence of any surface flux. The hydrogen diffusion equations, which are analogous to those for temperature diffusion, are incorporated within both mechanical model user material (UMAT) subroutines via the user material heat transfer (UMATHT) subroutine in Abaqus, which is called at the beginning of each time increment. Further details were provided by Elmukashfi et al. \cite{ELMUKASHFI2020}. Crystal plasticity subroutine codes \cite{DEMIR2025} used in this work are made available online \cite{GITHUB}. Codes developed for this manuscript may be made available on request. 

\section{Results and model development} \label{results}

\subsection{Micromechanical redistribution of hydrogen} \label{results1}

\begin{figure}[htb]
  \centering
  \includegraphics[width=0.95\linewidth]{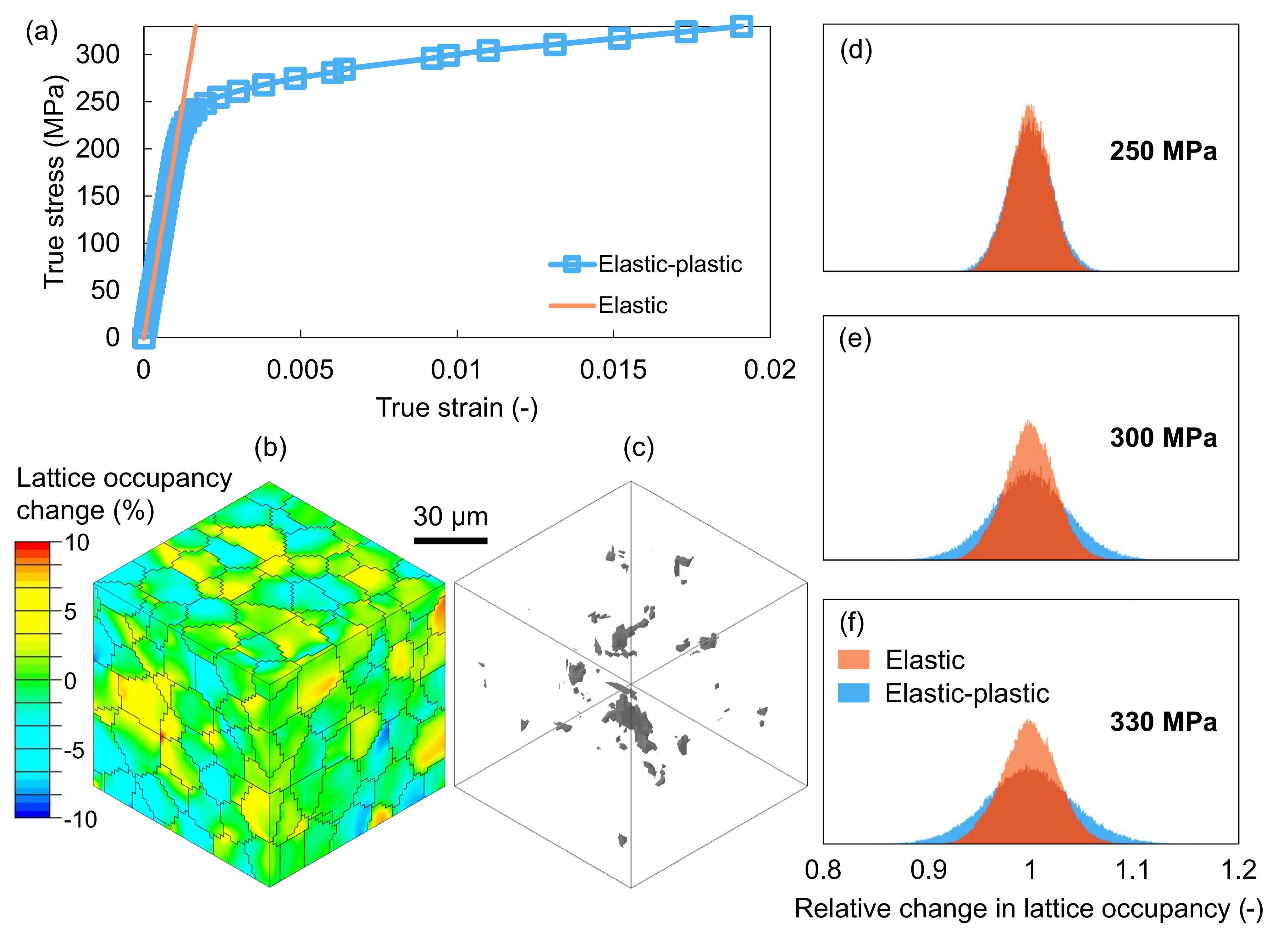}
  \caption{The relative effects of elastic and plastic anisotropy on the redistribution of hydrogen at saturation, i.e., as $t\to\infty$ within polycrystalline 316L. (a) Mechanical response of 316L with and without plasticity up to 330 MPa (equivalent to 2\% strain with plasticity). (b) Typical saturated distribution of hydrogen in untextured polycrystalline 316L after deformation; note that high and low values correspond to high and low values of hydrostatic stress, the locations of which are mostly near GBs and other stress concentration features. (c) As the distribution of hydrogen throughout the microstructure is found to conform well to a normal distribution, this subfigure shows a typical isosurface plot of regions within the microstructure which contain hydrogen concentrations beyond 3 standard deviations of the mean. (d) - (f) Frequency distributions of the relative change in lattice occupancy ($\theta_{\mathrm{L}}/\theta_{\mathrm{L}}^{\mathrm{ini}}$) at stresses 250, 300, and 330 MPa, respectively, with and without plasticity.}
  \label{figElasPlas}
\end{figure}

The steady state grain scale distributions of diffusible hydrogen in 316L are studied in this subsection using the micromechanical model presented in subsection \ref{MicroModel}, which is coupled with the hydrogen transport model. As steady state concentrations are independent of lattice diffusivity, $D_{\mathrm{L}}$, this is assigned a non-physical value which yields steady state results within reasonable computational time scales, unlike the diffusivity for 316L, which is extremely low \cite{BRASS2006}. For each micromechanical model, the initial lattice occupancy is distributed homogeneously and is prescribed a value of $10^{-3}$. At the boundaries, out-of plane fluxes are zero, such that the total hydrogen content remains unchanged during each simulation. The gradients of hydrostatic stress generated by microstructure anisotropy are known to drive hydrogen redistribution \cite{HUSSEIN2021}, which is captured in this framework by the associated chemical potential gradients. However, it is not yet understood how the local elastic constants (or indeed localised plasticity) contribute to these hydrostatic stress gradients, or ultimately how this is linked to the localisation of hydrogen. As an initial case study in 316L, the relative contributions of elastic and elastic-plastic anisotropy on the steady state redistribution of hydrogen are rationalised. To do this, as shown in Figure \ref{figElasPlas} (a), the full micromechanical model is utilised to generate a mechanical response up to around 2\% total strain corresponding to a mean stress of 330 MPa; for comparison against the equivalent elastic response, $\tau_{\mathrm{c},0}$ is set to a very high value (prohibiting the onset of slip), and the polycrystal is loaded to 330 MPa. A comparison of hydrogen distributions at regular intervals of stress can help to differentiate between elastic and plastic contributions. As anticipated \cite{HUSSEIN2021}, each lattice occupancy distribution is found to conform to the Gaussian distribution. A sample contour plot result showing the distribution of lattice occupancy throughout the microstructure is presented in Figure \ref{figElasPlas} (b). For the same distribution, an isosurface plot of the regions which lie beyond the 99.7\% confidence interval of the Gaussian distribution ($\pm$3 standard deviations of the mean) is shown in Figure \ref{figElasPlas} (c). The largest isosurface feature is found to measure less than half of the mean grain size, and hence values which lie beyond the 3 standard deviation limit are not considered here. Figure \ref{figElasPlas} (d) - (f) show the frequency distributions of the relative change in lattice occupancy ($\theta_{\mathrm{L}}/\theta_{\mathrm{L}}^{\mathrm{initial}}$) for fully elastic and elastic-plastic cases at 250, 300, and 330 MPa, respectively. At 250 MPa, the distributions are found to closely overlap as plasticity levels remain relatively low in the elastic-plastic case. At 300 MPa however, both distributions are shown to broaden due to the increase in stress, with the elastic-plastic case broadening substantially more. This trend is shown to continue at 330 MPa, with further broadening of each distribution, driving local hydrogen concentrations beyond 15\% higher than the initial value. 

\begin{figure}[htb]
  \centering
  \includegraphics[width=0.626\linewidth]{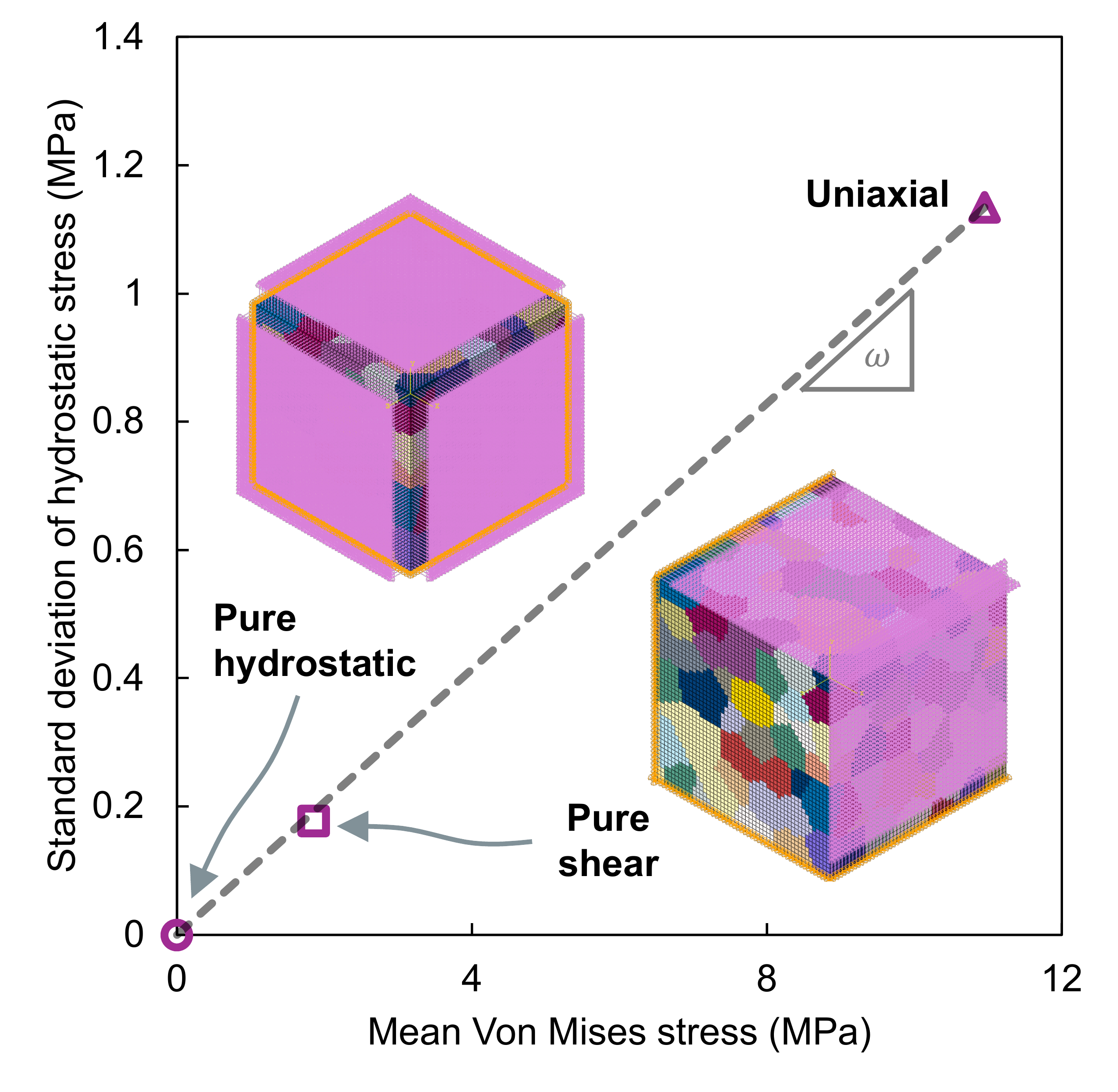}
  \caption{The relationship between the mean von Mises stress across an elastically deformed polycrystal with cubic crystallography and the standard deviation of the normally distributed hydrostatic stress. Extreme examples of pure hydrostatic, pure shear, and uniaxial loading are given. The linear relationship is characterised by the stress ratio, $\omega$.}
  \label{figOmega}
\end{figure}

Naturally, it will be essential to characterise the hydrostatic stress distribution before evaluating its effect on the localisation of hydrogen. Interestingly, hydrostatic stress distributions in 316L are also found here to conform to the Gaussian model, which was not reported in previous microstructural hydrogen distribution studies \cite{HASSAN2019,ABDOLVAND2019,HUSSEIN2021}. However, in a statistical analysis of intergranular normal stresses (INSs), El Shawish \cite{ELSHAWISH2024} noted that in linear elastic loading of cubic polycrystals, INS (which is analogous to the hydrostatic stress) is normally distributed, and its variance was shown to be proportional to the $J_2$ deviatoric stress invariant. In other words, the standard deviation of the INS distribution is proportional to the mean von Mises stress. In orthorhombic and other crystal structures with non-cubic symmetry (which are not considered here), the INS distributions were shown to also depend on the mean hydrostatic stress. Similar findings were reported in an early CPFE modelling study by Kozaczek et al. \cite{KOZACZEK1995} and in a statistical analysis by Fokin and Shermergor \cite{FOKIN1968}. To evaluate this in the context of 316L, three extreme cases of RVE elastic loading are considered: pure hydrostatic loading, pure shear loading, and uniaxial loading (which represents a combination of the other two). As illustrated in Figure \ref{figOmega}, these three cases are found to comply with this law of proportionality, for the case of pure volumetric loading ($\sigma_{\mathrm{H}}=1$ MPa) no internal hydrostatic stress variation is observed, as represented by the standard deviation, $\Delta\sigma_{\mathrm{H}}$. $\Delta\sigma_{\mathrm{H}}$ is obtained by fitting to the Gaussian distribution function. The dimensionless parameter, $\omega$, is introduced in Figure \ref{figOmega} to characterise the the ratio, $\Delta\sigma_{\mathrm{H}}/\Bar{\sigma}_{\mathrm{VM}}$, for different materials, where $\Bar{\sigma}_{\mathrm{VM}}$ is the mean von Mises stress. 

It is understandable for polycrystals with cubic symmetry that by imposing a volumetric strain or stress, there is no internal variation in stress, since volume changes are isotropic and invariant to crystal orientation. The elastic anisotropy in 316L arises from the shear terms of stiffness, $\frac{C_{11}-C_{12}}{2}$ and $C_{44}$, which represent the resistance to shear along the $<110>$ and $<100>$ directions, respectively \cite{KNOWLES2015}. The level of cubic anisotropy is often characterised by the Zener anisotropy ratio, $A=\frac{2C_{44}}{C_{11}-C_{12}}$. Hence, it follows that for 316L and other anisotropic cubic polycrystals, internal hydrostatic stress variations should vary according to the applied deviatoric stress only. This is discussed in greater detail in the next subsection. 

\subsection{Analytical sub-modelling of hydrostatic stresses and strain-rate dependence}

In this subsection, rate-dependent microscale diffusion is considered. Analytical models are developed based on results and understanding obtained from CPFE modelling, and are later used for implementation at the continuum level in subsection \ref{notched_modelling}. According to Fokin and Shermergor \cite{FOKIN1968} and in a summary of their work by Dikusar et al. \cite{DIKUSAR1971}, the standard deviation of the 'internal stresses' in an elastically deformed polycrystal is proportional to the 'applied stress'. Based on findings from subsection \ref{results1}, the analytical model derived by Fokin and Shermergor \cite{FOKIN1968} is considered here in evaluating the relationship between $\Delta\sigma_{\mathrm{H}}$ and $\Bar{\sigma}_{\mathrm{VM}}$, giving,

\begin{equation}
\Delta\sigma_{\mathrm{H}}=\omega{\Bar{\sigma}}_{\mathrm{VM}} \label{SD}
\end{equation}

where $\omega$ is obtained from \cite{FOKIN1968} and is given by,

\begin{equation}
\omega=\frac{0.145K|\lambda|}{(K+\frac{4}{3}G)E} \label{omega}
\end{equation}

where $K$ represents the bulk modulus, $G$ the shear modulus, $E$ Young's modulus, and $\lambda$ represents the degree of mismatch between the $<110>$ and $<100>$ elastic shear terms (which were discussed earlier), and is given by Equation (\ref{lambda}).

\begin{equation}
\lambda=C_{11}-C_{12}-2C_{44} \label{lambda}
\end{equation}

\begin{figure}[htb]
  \centering
  \includegraphics[width=0.95\linewidth]{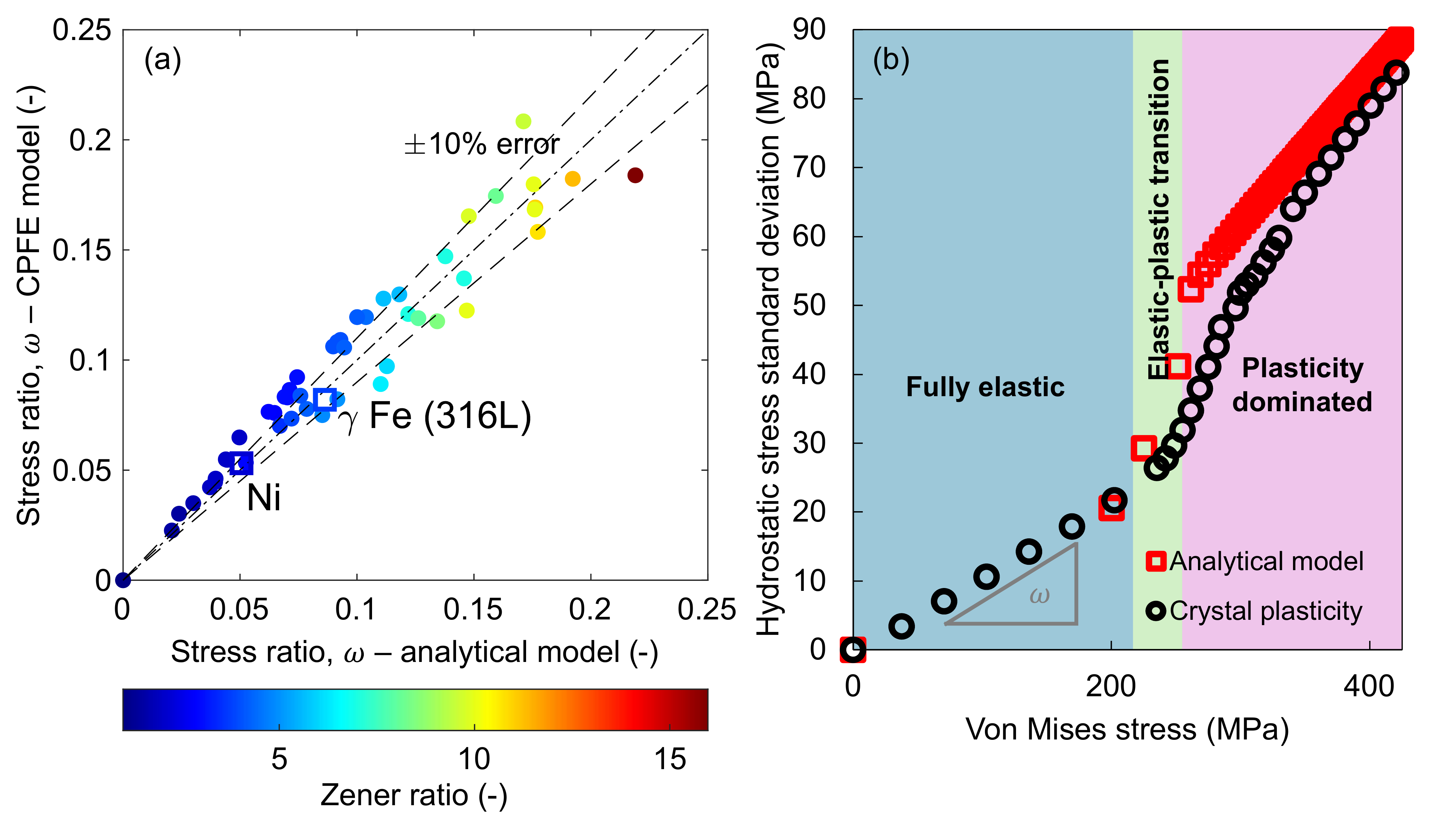}
  \caption{(a) Comparison of the predicted stress ratio for randomly textured cubic polycrystals (for unique combinations of elastic constants, $C_{11}$, $C_{12}$, and $C_{44}$) with CPFE-predicted values. Values specific to nickel and 316L are also included. (b) Comparison of analytical and CPFE models in capturing the evolution of the hydrostatic stress distribution with increasing mean von Mises stress in fully elastic, transitionary, and plasticity-dominated regimes.}
  \label{figRatio}
\end{figure}

To gauge this model against CPFE-predicted results, 50 random combinations of elastic constants are generated in the range, $C_{11}\in[100,300]$, $C_{12}\in[50,150]$, and $C_{44}\in[50,150]$ GPa, and implemented within the 291 grain micromechanical model for extraction of statistical data. Upon application of a known mean von Mises stress (below the yield stress), the standard deviation of hydrostatic stress was obtained for each combination by fitting the data to the Gaussian distribution function, as discussed earlier. The dimensionless stress ratio was then simply obtained via Equation (\ref{SD}). For the analytical model, the same elastic constant combinations were applied, leaving $K$, $G$, and $E$ to be determined for each case. For each combination, the stiffness matrix $\boldsymbol{C}$, was generated and rotated 1000 times (at random, across the full rotation spectrum) to give a statistically significant representation of orientations. $E$ and $G$ were approximated as $E=1/\bar{S}_{11}$ and $G=1/\bar{S}_{44}$, respectively, from the mean of the distribution of $N=1000$ compliance matrices, $\boldsymbol{S}=\inv{\boldsymbol{C}}$, e.g., $\bar{S}_{11}=\frac{1}{1000}\sum_{n=1}^{N=1000}\boldsymbol{S}_n(1,1)$. The bulk modulus, $K$, was obtained for each case from $K=E/{3(1-2\nu)}$, where $\nu$ is Poisson's ratio and is obtained from the known relationship between shear and Young's moduli, $G=E/{2(1+\nu)}$. The results of this study are presented in Figure \ref{figRatio} (a), showing reasonably strong agreement between the analytical and numerical modelling methods. Marker points are colour coded according to Zener anisotropy ratio, which was found correlate with $\omega$, though rather weakly. Predicted $\omega$ values for 316L and pure nickel (which are used as case study materials in subsection \ref{notched_modelling}) are also presented in Figure \ref{figRatio} (a) using square markers. 

To incorporate the effect of plastic deformation on $\omega$, an adaptation is made to Equation (\ref{omega}), accounting for the stiffness degradation associated with plastic shear. Hence, the tangent stiffness degradation factor, $\chi\in[0,1]$ is introduced here as,

\begin{equation}
\chi=\frac{1}{E}\frac{\partial\sigma_{\mathrm{y}}}{\partial\epsilon_{\mathrm{VM}}} \label{chi}
\end{equation}

where $\frac{\partial\sigma_{\mathrm{y}}}{\partial\epsilon_{\mathrm{VM}}}$ represents the tangent stiffness (with respect to total strain) of the macroscopic stress-strain response. Here, the von Mises strain, $\epsilon_{\mathrm{VM}}$, is used to represent the total strain under multiaxial loading. Hence, assuming tangent stiffness degradation is facilitated by plastic shear only, this is accounted for in the plasticity-corrected stress ratio, $\omega^{\mathrm{p}}$, by scaling the shear modulus, $G$, with the degradation factor, $\chi$, as follows,

\begin{equation}
\omega^{\mathrm{p}}=\frac{0.145K|\lambda^{\mathrm{p}}|}{(K+\frac{4}{3}G\chi)E} \label{omegaPC}
\end{equation}

where $\lambda^{\mathrm{p}}$ is the plasticity-corrected $\lambda$, and is obtained by accounting for the local degradation of the $\frac{1}{2}(C_{11}-C_{12})$ stiffness term (see Equation (\ref{lambdaPC})), since it represents the resistance to shear along the $<110>$ direction, i.e., the assumed direction of slip for FCC metals. 

\begin{equation}
\lambda^{\mathrm{p}}=(C_{11}-C_{12})\chi-2C_{44} \label{lambdaPC}
\end{equation}

In Equation (\ref{omegaPC}), it is assumed that the degradation of tangent stiffness is attributed directly to relaxation of the bulk and local shear resistance terms, $G$, and $\frac{1}{2}(C_{11}-C_{12})$. The bulk modulus, $K$, which relates hydrostatic stress to hydrostatic strain, is assumed unchanged since at the continuum level, hydrostatic stress does not influence slip and vice versa. In any case, as $\chi\to0$, Equations (\ref{omegaPC}) and (\ref{lambdaPC}) combine to give Equation (\ref{omegaPC_inf}) for the dimensionless gradient at plastic saturation, i.e., no hardening, $\omega^{\mathrm{p}}_{\infty}$, which is independent of $K$. 

\begin{equation}
\omega^{\mathrm{p}}_{\infty}=\frac{0.29C_{44}}{E} \label{omegaPC_inf}
\end{equation}

During the elastic-plastic transition, the rise in von Mises stress is assumed to have no contribution to the elastic interactions between neighbouring grains, since internal stresses become completely redistributed due to the addition of local plastic anisotropy. As plasticity percolates the microstructure, the hydrostatic stresses generated due to elastic anisotropy are therefore redistributed until the material enters the plasticity-dominated regime, which is illustrated in Figure \ref{figRatio} (b). Hence, the three stages of deformation considered here (fully elastic, elastic-plastic transition, and plasticity-dominated), are captured within this analytical framework as follows,

\begin{equation}\label{eqEPtrans}
    \Delta\sigma_{\mathrm{H}}= 
\begin{cases}
    \omega\bar{\sigma}_{\mathrm{VM}},& \text{if } \bar{\sigma}_{\mathrm{VM}}\leq\sigma_{\mathrm{y}}^0\\
    \omega^{\mathrm{p}}\sigma_{\mathrm{y}}^0,& \text{if } \sigma_{\mathrm{y}}^0<\bar{\sigma}_{\mathrm{VM}}<\sigma_{\mathrm{y}}^{\mathrm{p}}\\
    \omega^{\mathrm{p}}(\bar{\sigma}_{\mathrm{VM}}-\sigma_{\mathrm{ep}}),& \text{otherwise}
\end{cases}
\end{equation}

where $\sigma_{\mathrm{y}}^{\mathrm{p}}$ is the stress at the end of the elastic-plastic transition (corresponding to the initial stage at which there is plastic flow in every grain), and $\sigma_{\mathrm{ep}}=\sigma_{\mathrm{y}}^{\mathrm{p}}-\sigma_{\mathrm{y}}^0$ is the stress range covered during the elastic-plastic transition. This analytical formulation for predicting the distribution of hydrostatic stress in cubic polycrystals is compared against the CPFE modelling results for 316L in Figure \ref{figRatio} (b), showing overall reasonable agreement. Interestingly, there is a sharp rise in $\Delta\sigma_{\mathrm{H}}$ during the elastic-plastic transition due to the sudden tangent stiffness degradation. This is followed by a near linear evolution of $\Delta\sigma_{\mathrm{H}}$ in the plasticity-dominated regime, at a steeper gradient ($\omega^{\mathrm{p}}$) than that for fully elastic loading ($\omega$). Hence, this illustrates the importance of considering plastic anisotropy in studying the local redistribution of hydrogen. 

Now that the standard deviation of hydrostatic stress is attainable for cubic polycrystals undergoing elastic-plastic monotonic loads, additional analytical models for the transient evolution of localised hydrogen may be developed. As outlined in Equation (\ref{dcldt}), the rate of change of local lattice concentration is proportional to the spatial gradient of local flux, $J^{\mathrm{loc}}$, expressed here as a scalar quantity to represent the mean change in any direction within a polycrystal. Here, a linear approximation is used in which some flux change, $\delta J^{\mathrm{loc}}$, is assumed over the characteristic length scale for diffusion, $l_{\mathrm{c}}$. It is assumed that in the absence of any external or surface hydrogen flux, diffusion is confined to the characteristic length scale (grain scale), giving mean flux $\bar{J}=0$ so that $\delta J^{\mathrm{loc}}=\pm J^{\mathrm{loc}}$, depending on whether there is a local increase or decrease in hydrogen content. Equation (\ref{dcldt}) is formulated to give an initial local increase in hydrogen, as in subsequent expressions, the positive hydrostatic stress standard deviation, $\Delta\sigma_{\mathrm{H}}$, is used to characterise the localisation of stress at the grain scale. 

\begin{equation}
\frac{\partial C_{\mathrm{L}}^{\mathrm{loc}}}{\partial t}=-\frac{\partial J^{\mathrm{loc}}}{\partial x}=-\frac{1}{l_{\mathrm{c}}}(0-J^{\mathrm{loc}})\label{dcldt}
\end{equation}

As in Equation (\ref{dcldt}), the chemical potential gradient is also approximated as linear in Equation (\ref{dmudx1}), by considering the local change in chemical potential, $\delta\mu=\mu^{\mathrm{loc}}-\bar{\mu}$, where $\mu^{\mathrm{loc}}$ represents a local chemical potential concentration, and $\bar{\mu}$ is the mean chemical potential. From Equation (\ref{mu}), (for small $C_{\mathrm{L}}$, lattice occupancy, $\theta_{\mathrm{L}}=\frac{C_{\mathrm{L}}}{C_{\mathrm{L}}^{\mathrm{max}}}\approx\frac{C_{\mathrm{L}}}{C_{\mathrm{L}}^{\mathrm{max}}-C_{\mathrm{L}}}$) the chemical potential gradient is then,

\begin{equation}
\frac{\partial\mu^{\mathrm{loc}}}{\partial x}=\frac{1}{l_{\mathrm{c}}}(\mu^{\mathrm{loc}}-\bar{\mu})=\frac{1}{l_{\mathrm{c}}}\left((RT\ln{\theta_{\mathrm{L}}^{\mathrm{loc}}}+\mu_{\sigma_{\mathrm{H}}}^{\mathrm{loc}})-(RT\ln{\bar{\theta}_{\mathrm{L}}}+\bar{\mu}_{\sigma_{\mathrm{H}}})\right) \label{dmudx1}
\end{equation}

where $\theta_{\mathrm{L}}^{\mathrm{loc}}$ and $\mu_{\sigma_{\mathrm{H}}}^{\mathrm{loc}}$ represent the local lattice occupancy and local contribution to chemical potential by hydrostatic stress and $\bar{\theta}_{\mathrm{L}}$ and $\bar{\mu}_{\sigma_{\mathrm{H}}}$ represent the mean lattice occupancy and contribution to chemical potential by hydrostatic stress, respectively. Letting the hydrostatic stress standard deviation represent the difference, $\Delta\sigma_{\mathrm{H}}=\sigma_{\mathrm{H}}^{\mathrm{loc}}-\bar{\sigma}_{\mathrm{H}}$ gives Equation (\ref{dmudx2}) as follows, 

\begin{equation}
\frac{\partial\mu^{\mathrm{loc}}}{\partial x}=\frac{1}{l_{\mathrm{c}}}\left(RT\ln{\frac{\theta_{\mathrm{L}}^{\mathrm{loc}}}{\bar{\theta}_{\mathrm{L}}}}-\Delta\sigma_{\mathrm{H}}V_{\mathrm{L}}\right) \label{dmudx2}
\end{equation}

where $\bar{\mu}_{\sigma_{\mathrm{H}}}=\bar{\sigma}_{\mathrm{H}}V_{\mathrm{L}}$ and $\mu_{\sigma_{\mathrm{H}}}^{\mathrm{loc}}=\sigma_{\mathrm{H}}^{\mathrm{loc}}V_{\mathrm{L}}$. Combining Equations (\ref{dcldt}) and (\ref{dmudx2}) yields Equation (\ref{dcldt2}) for the rate dependent evolution of local lattice concentration. 

\begin{equation}
\frac{\partial C_{\mathrm{L}}^{\mathrm{loc}}}{\partial t}=-\frac{D_{\mathrm{L}}C_{\mathrm{L}}^{\mathrm{loc}}}{RTl_{\mathrm{c}}}\nabla\mu=-\frac{D_{\mathrm{L}}C_{\mathrm{L}}^{\mathrm{loc}}}{l_{\mathrm{c}}^2}\left(\ln{\frac{\theta_{\mathrm{L}}^{\mathrm{loc}}}{\bar{\theta}_{\mathrm{L}}}}-\frac{\Delta\sigma_{\mathrm{H}}V_{\mathrm{L}}}{RT}\right) \label{dcldt2}
\end{equation}

Where the mean lattice occupancy is effectively unchanged (assuming low trap density evolution at low levels of strain), the local concentration, $C_{\mathrm{L}}^{\mathrm{loc}}$, is proportional to the lattice occupancy ratio, $\theta_{\mathrm{L}}^{\mathrm{loc}}/\bar{\theta}_{\mathrm{L}}$. Hence, Equation (\ref{dcldt2}) is rewritten to give Equation (\ref{dthetaldt}).

\begin{equation}
\frac{\partial\theta_{\mathrm{L}}^{\mathrm{loc}}/\bar{\theta}_{\mathrm{L}}}{\partial t}=-\frac{D_{\mathrm{L}}}{l_{\mathrm{c}}^2}\frac{\theta_{\mathrm{L}}^{\mathrm{loc}}}{\bar{\theta}_{\mathrm{L}}}\left(\ln{\frac{\theta_{\mathrm{L}}^{\mathrm{loc}}}{\bar{\theta}_{\mathrm{L}}}}-\frac{\Delta\sigma_{\mathrm{H}}V_{\mathrm{L}}}{RT}\right) \label{dthetaldt}
\end{equation}

Notice that Equation (\ref{dthetaldt}) now takes the form of a nonlinear differential equation. Defining the characteristic diffusion time, $\tau=\frac{l_{\mathrm{c}}^2}{D_{\mathrm{L}}}$, and imposing the initial condition at $t=0$, $\theta_{\mathrm{L}}^{\mathrm{loc}}/\bar{\theta}_{\mathrm{L}}=1$ (since prior to diffusion, hydrogen is distributed homogeneously), we obtain the solution given by Equation (\ref{theta_l^loc}).

\begin{equation}
\frac{\theta_{\mathrm{L}}^{\mathrm{loc}}}{\bar{\theta}_{\mathrm{L}}}(t)=\exp{\left(\frac{\Delta\sigma_{\mathrm{H}}V_{\mathrm{L}}}{RT}\left(1-\exp{\left(-\frac{t}{\tau}\right)}\right)\right)} \label{theta_l^loc}
\end{equation}

Since the lattice occupancy standard deviation is $\Delta\theta_{\mathrm{L}}=\theta_{\mathrm{L}}^{\mathrm{loc}}-\bar{\theta}_{\mathrm{L}}$, its rate dependent form is now easily derived from Equation (\ref{theta_l^loc}) to give Equation (\ref{deltathetaL}).  

\begin{equation}
\Delta\theta_{\mathrm{L}}=\bar{\theta}_{\mathrm{L}}\left(\exp{\left(\frac{\Delta\sigma_{\mathrm{H}}V_{\mathrm{L}}}{RT}\left(1-\exp{\left(-\frac{t}{\tau}\right)}\right)\right)}-1\right) \label{deltathetaL}
\end{equation}

Using the coarse linear approximation for flux and chemical potential gradients described here, it is necessary to account for the variation in characteristic length scale, $l_{\mathrm{c}}$, as gradients are initially dominated by the hydrostatic stress field, and assumed to later smoothen so that grain size becomes a good approximation for $l_{\mathrm{c}}$. Hence, the mean (1D) distribution of hydrostatic stress across a grain may be written as a function of distance from a stress concentration feature, $x$, in the classical way, $\sigma_{\mathrm{H}}=\alpha x^{-\frac{1}{2}}$, where $\alpha$ is some constant, and $x$ is distance from a GB stress concentration feature. The mean hydrostatic stress across a grain of size $d_{\mathrm{g}}$ is then obtained by integrating this function to give $\bar{\sigma}_{\mathrm{H}}=\frac{1}{d_{\mathrm{g}}}\int_{0}^{d_{\mathrm{g}}}\alpha x^{-\frac{1}{2}}dx=2\alpha x^{-\frac{3}{2}}$. The constant $\alpha$ then becomes $\frac{1}{2}\bar{\sigma}_{\mathrm{H}}d_{\mathrm{g}}^{\frac{1}{2}}$. The hydrostatic stress gradient may then also be defined and is given by $\frac{\partial\sigma_{\mathrm{H}}}{\partial x}=-\frac{1}{2}\alpha x^{-\frac{3}{2}}$. By taking this gradient and inverting it, the characteristic length scale for the hydrostatic stress field may be approximated by its product with the hydrostatic stress standard deviation. In Equation (\ref{l_c^sigma_1}), the gradient is taken at the upper end of the normal distribution ($\sigma_{\mathrm{H}}=\bar{\sigma}_{\mathrm{H}}+3\Delta\sigma_{\mathrm{H}}$) as higher hydrostatic stresses are assumed to dominate diffusion initially. 

\begin{equation}
l_{\mathrm{c}}^{\sigma_{\mathrm{H}}}=\Delta\sigma_{\mathrm{H}}\left|\left(\frac{\partial\sigma_{\mathrm{H}}}{\partial x}\right)^{-1}_{\bar{\sigma}_{\mathrm{H}+3\Delta\sigma_{\mathrm{H}}}}\right|=\Delta\sigma_{\mathrm{H}}\left|\left(-2\alpha^{-1}x^{\frac{3}{2}}\right)_{\bar{\sigma}_{\mathrm{H}+3\Delta\sigma_{\mathrm{H}}}}\right| \label{l_c^sigma_1}
\end{equation}

By writing $x$ in terms of hydrostatic stress, we have $x=\left(\frac{\alpha}{\sigma_{\mathrm{H}}}\right)^2$, which when combined with the earlier expression for $\alpha$ at $\sigma_{\mathrm{H}}=\bar{\sigma}_{\mathrm{H}}+3\Delta\sigma_{\mathrm{H}}$, yields Equation (\ref{l_c^sigma_2}).

\begin{equation}
l_{\mathrm{c}}^{\sigma_{\mathrm{H}}}=\frac{\bar{\sigma}_{\mathrm{H}}^2\Delta\sigma_{\mathrm{H}}}{2\left(\bar{\sigma}_{\mathrm{H}}+3\Delta\sigma_{\mathrm{H}}\right)^3}d_{\mathrm{g}} \label{l_c^sigma_2}
\end{equation}

\begin{figure}[htb]
  \centering
  \includegraphics[width=0.96\linewidth]{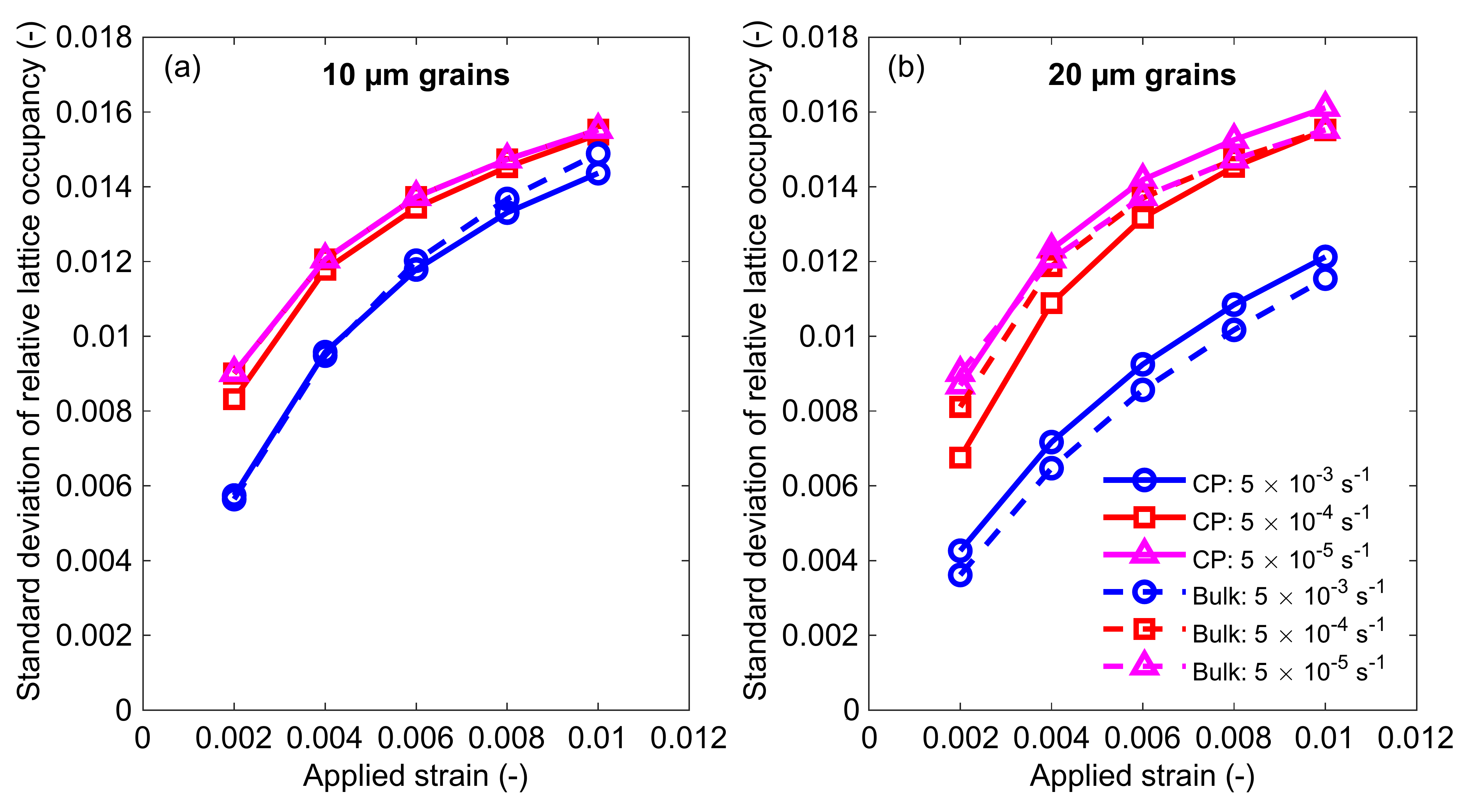}
  \caption{Capturing the strain-rate sensitivity of relative lattice occupancy distributions for a material with arbitrary mechanical and hydrogen transport properties using analytical modelling. Results for polycrystals with mean grain sizes of (a) 10 $\upmu$m and (b) 20 $\upmu$m are shown. Legend labels 'CP' and 'Bulk' correspond to detailed micromechanical (crystal plasticity) modelling results (considered ground truth) and microstructure-enriched continuum modelling results, respectively.} \label{figratesensitivity}
\end{figure}

A simple weighted mean function is then employed to scale the absolute contributions of the hydrostatic stress ($\Delta\sigma_{\mathrm{H}}V_{\mathrm{L}}$) and redistributed hydrogen ($RT\ln{{\theta_{\mathrm{L}}^{\mathrm{loc}}}/{\bar{\theta}_{\mathrm{L}}}}$) to the chemical potential gradient, $\frac{\partial\mu^{\mathrm{loc}}}{\partial x}$, to obtain a characteristic length scale which evolves with the redistribution of hydrogen as follows. In Figure \ref{figratesensitivity}, the results of this analytical framework are compared against CPFE modelling results for microstructures with mean grain sizes of 10 $\upmu$m and 20 $\upmu$m, subjected to monotonic and uniaxial loading as before, now at three distinct strain-rates which drive differing levels of hydrogen redistribution: $5\times10^{-3}$ s$^{-1}$, $5\times10^{-4}$ s$^{-1}$, and $5\times10^{-5}$ s$^{-1}$. 

\begin{equation}
l_{\mathrm{c}}=l_{\mathrm{c}}^{\sigma_{\mathrm{H}}}\left(\frac{\Delta\sigma_{\mathrm{H}}V_{\mathrm{L}}}{RT\ln{\frac{\theta_{\mathrm{L}}^{\mathrm{loc}}}{\bar{\theta}_{\mathrm{L}}}}+\Delta\sigma_{\mathrm{H}}V_{\mathrm{L}}}\right) + d_{\mathrm{g}}\left(\frac{RT\ln{\frac{\theta_{\mathrm{L}}^{\mathrm{loc}}}{\bar{\theta}_{\mathrm{L}}}}}{RT\ln{\frac{\theta_{\mathrm{L}}^{\mathrm{loc}}}{\bar{\theta}_{\mathrm{L}}}}+\Delta\sigma_{\mathrm{H}}V_{\mathrm{L}}}\right)
\label{l_c}
\end{equation}

Figure \ref{figratesensitivity} presents the predicted relative microstructure-driven hydrogen redistribution, $\theta_{\mathrm{L}}^{\mathrm{loc}}/\bar{\theta}_{\mathrm{L}}-1$, at various levels of deformation between 0.2\% and 1.0\% total strain for each microstructure and strain-rate. The micromechanical properties are for 316L, while the chosen diffusivity is arbitrary, as before. The CPFE results show that with an increase in applied strain (corresponding to an increase in the mean von Mises stress and plastic strain level), the localisation of hydrogen increases nonlinearly. It is also clear for both grain sizes that lower strain-rates yield larger levels of redistribution. Both grain size cases appear to approach saturation at the lowest strain-rate, with greatest overall differences between strain-rates in the 20 $\upmu$m case. This is due to lower overall spatial gradients of chemical potential in the 20 $\upmu$m case (assuming distributions of stress are independent of grain size in the absence of GNDs). The analytical model, labelled in Figure \ref{figratesensitivity} as 'Bulk', is shown to agree reasonably well with the CPFE model, particularly for the 10 $\upmu$m case, though is perhaps slightly conservative in approaching saturation nearer the intermediate strain-rate. 

\subsection{Multiscale modelling of notched tensile test specimens} \label{notched_modelling}

\begin{figure}[htb]
  \centering
  \includegraphics[width=1\linewidth]{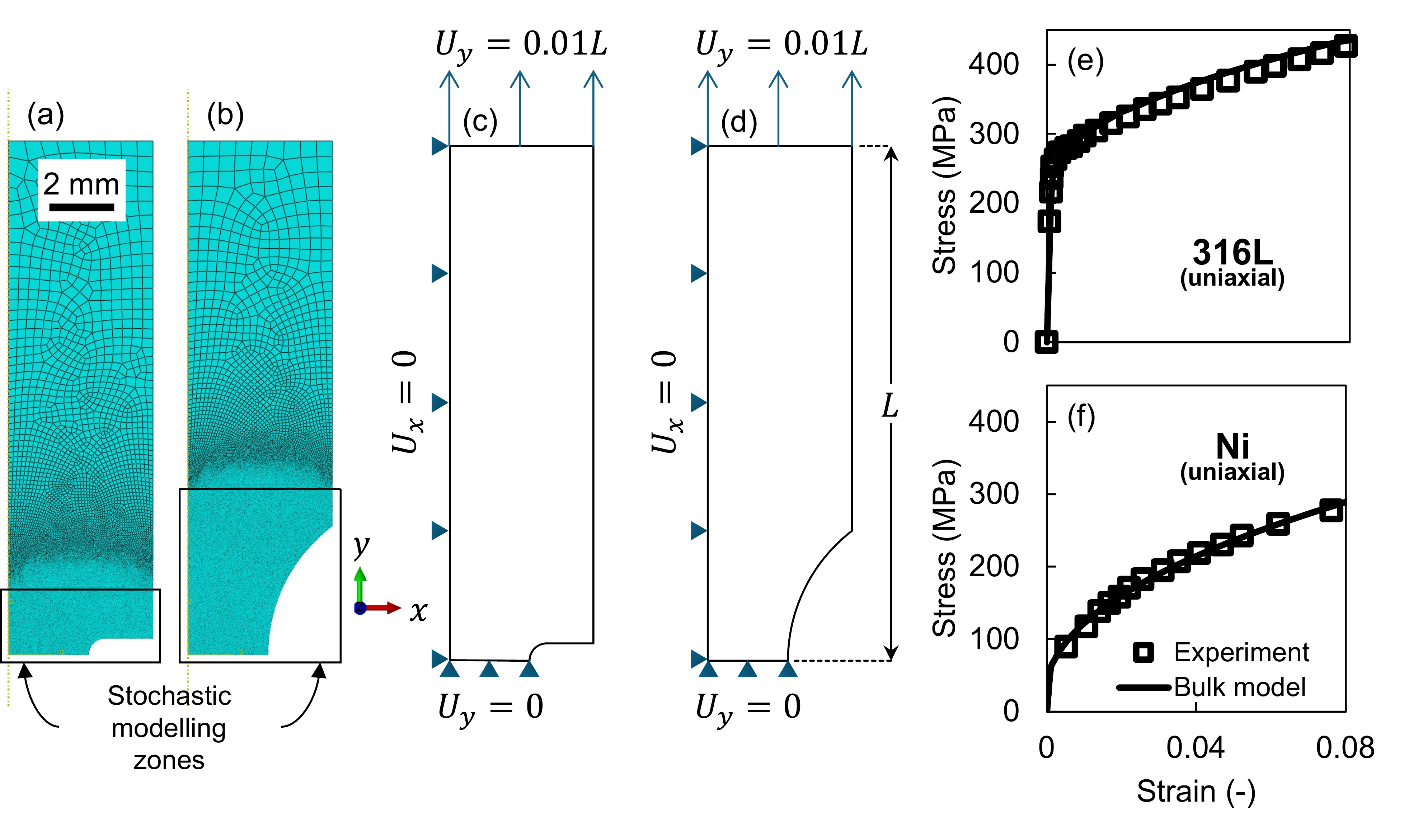}
  \caption{Details of mesh used in axisymmetric modelling of notched tensile test specimens are given for notch radii (a) 0.5 mm and (b) 5 mm. Stochastic modelling zones in which the statistical distributions of hydrostatic stress and hydrogen concentration are considered are outlined. Boundary and loading conditions for these respective geometries are given in (c) and (d). The displacement, $U_y$, is applied over various time periods to give different mean axial strain-rates. Subfigures (e) and (f) show the experimental \cite{KANG2010,ARYA2021} and modelled (mm-scale) mechanical responses of 316L and pure nickel, respectively.} 
  \label{axisymmetric}
\end{figure}

In this subsection, the continuum mechanical model (subsection \ref{MacroModel}) is enriched by incorporating the characteristics of microscale hydrogen redistribution. Mechanistic understanding gained from simulating microstructural hydrogen redistribution is incorporated analytically to capture relevant statistical information within millimetre scale mechanical and hydrogen diffusion models. Two notched axisymmetric tensile test geometries are considered, with notch radii of 0.5 mm and 5 mm, shown in Figure \ref{axisymmetric} (a) and (b), respectively. 

Minimum and maximum sample diameters are 5 mm (at the notch root) and 9 mm (along the cylindrical section) in each case. Figure \ref{axisymmetric} (a) and (b) also introduce stochastic modelling zones, within which the statistical information from microstructural sub-models are incorporated. Specifically, the standard deviations of hydrostatic stress and lattice occupancy are calculated from Equations (\ref{eqEPtrans}) and (\ref{deltathetaL}) at each material point, and are used to estimate a spatially representative distribution of hydrostatic stress and lattice occupancy within the stochastic modelling zones based on a pre-assigned Z-score, $Z$, which conforms to the normal distribution. Specifically, elements within the stochastic modelling zone are set to the size of the mean grain size (here, $d_{\mathrm{g}}=10$ $\upmu$m), and each element is assigned a value at random for $Z\in[-3,3]$, representing the normal distribution in terms of standard deviations from the mean. The result is a representative distribution in terms of shape (Gaussian) and length scale. From there, representative distributions of hydrostatic stress and lattice occupancy are obtained from Equations (\ref{bulk_loc_sigH}) and (\ref{bulk_loc_thetaL}). 

\begin{subequations}
\begin{equation}
\sigma_{\mathrm{H}}^{\mathrm{loc}}=\bar{\sigma}_{\mathrm{H}}+Z\Delta\sigma_{\mathrm{H}} \label{bulk_loc_sigH}
\end{equation}    
\begin{equation}
\theta_{\mathrm{L}}^{\mathrm{loc}}=\bar{\theta}_{\mathrm{L}}+Z\Delta\theta_{\mathrm{L}} \label{bulk_loc_thetaL}
\end{equation}
\end{subequations}

\begin{figure}[htb]
  \centering
  \includegraphics[width=1\linewidth]{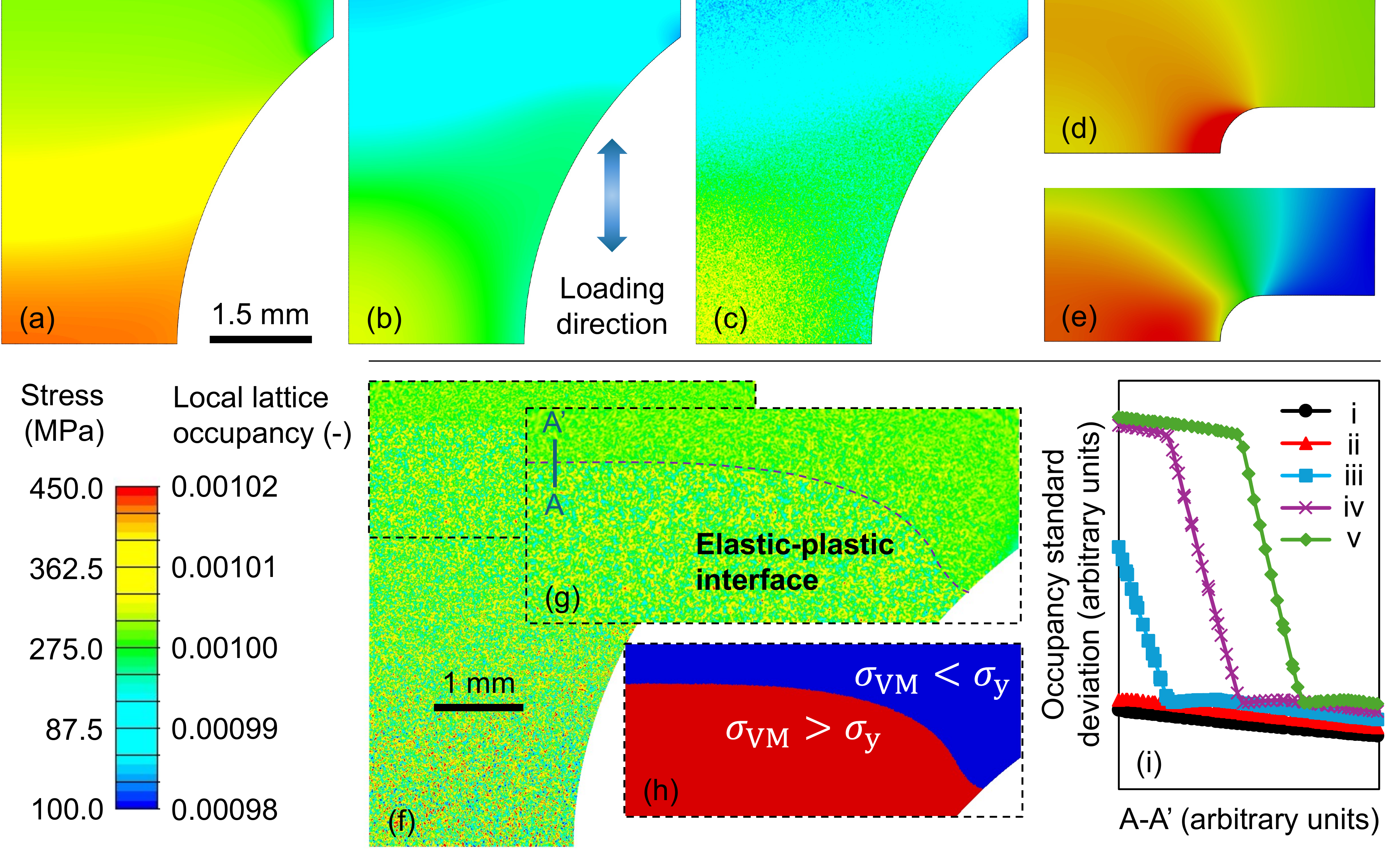}
  \caption{Modelled distributions of (a) macroscopic von Mises stress, (b) macroscopic hydrostatic stress, and (c) microstructure-level hydrostatic stress in 316L (5 mm notch radius) after at an equivalent applied strain of 1\%. The corresponding von Mises and hydrostatic stress distributions are shown for the sample with 0.5 mm notch radius in (d) and (e), respectively. The loading direction shown corresponds to all subfigures. (f) Microstructure-saturated distribution of local lattice occupancy in 316L sample with 5 mm notch radius. Subfigure (g) highlights the stark transition in lattice occupancy heterogeneity across the elastic-plastic interface, the meaning of which is illustrated in (h). (i) Distribution of lattice occupancy standard deviation along path A-A' from (g). The line A-A' is fixed in space and lines i - v correspond to chronological points in time. At i, deformation is fully elastic. Plasticity is shown to propagate through the path between iii and v.}
  \label{EPinterface}
\end{figure}

Monotonic displacement-controlled loads are applied as shown in Figure \ref{axisymmetric} (c) and (d) to 1\% mean strain along the axis of symmetry ($\bar{\epsilon}_{\mathrm{ax}}=U_{\mathrm{y}}(t)/L$). A wide range of strain-rates, which are defined according to this measure of strain, are considered in order to capture the full range over which microstructure and notch geometry influence diffusion in 316L and nickel. Hence, the experimental mechanical responses \cite{KANG2010,ARYA2021} for both materials are shown and compared against the bulk mechanical model shown in Figure \ref{axisymmetric} (e) and (f). The multiscale modelling properties which are used in this subsection are given in Table \ref{MultiProps}. Resultant distributions of stress and local lattice occupancy in 316L are presented in Figure \ref{EPinterface}.

For the blunt notch, the von Mises stress is shown to localise near the notch root and uniformly along the radial direction in Figure \ref{EPinterface} (a), as compared to hydrostatic stress (mean), which is concentrated towards the centre of the specimen in Figure \ref{EPinterface} (b). The local microstructure-affected hydrostatic stress distribution shown in Figure \ref{EPinterface} (c) highlights the seemingly unsubstantial variation relative to its continuum level counterpart, i.e., the greatest hydrostatic stress variation occurs over the notch length scale. Figures \ref{EPinterface} (d) and (e) show, respectively, the mean von Mises and hydrostatic stress distributions manifested by the sharp notch geometry. Interestingly, Figure \ref{EPinterface} (f) shows that despite the feeble contribution of microstructure to the overall local hydrostatic stress, local hydrogen concentrations are driven in some instances by up to 10\% beyond the initial value (at steady state microstructure diffusion), whilst notch driven diffusion is negligible by comparison. The equivalent is observed for the sharp notch. Figures \ref{EPinterface} (g) and (h) highlight that along the elastic-plastic interface, there is a very clear difference in the extent to which hydrogen is localised: there is much greater hydrogen content variation in the plastic zone than in the elastic zone. This reinforces the notion that plasticity is a very important consideration in anisotropy-driven hydrogen diffusion analyses. Figures \ref{EPinterface} (c), (f), and (g) present the local hydrostatic stress and lattice occupancy distributions by superposition of continuum scale fields with statistical representations of local microstructural quantities. This novel approach enables the reader to visualise the influence of microstructure relative to that of continuum morphological features in the context of hydrogen localisation and the spatial variation thereof. An important feature this highlights, is that despite the seemingly minimal microstructure-driven hydrostatic stress variation, microstructure-driven lattice occupancy variation dominates over continuum diffusion, owning to the large chemical potential gradients generated at low length scales. Figure \ref{EPinterface} (i) shows the distribution of local lattice occupancy standard deviation (in arbitrary units) along the path A-A' at different points in the loading history; lines i - v correspond to remote applied strains of 0.6\% to 1.0\% in intervals of 0.1\%, respectively. The moving plastic front is shown to drive a substantial increase in hydrogen localisation, particularly when compared to the slight change driven by a 0.1\% strain increase in either fully elastic or plasticity dominated regimes.

\begin{figure}[t!]
  \centering
  \includegraphics[width=0.86\linewidth]{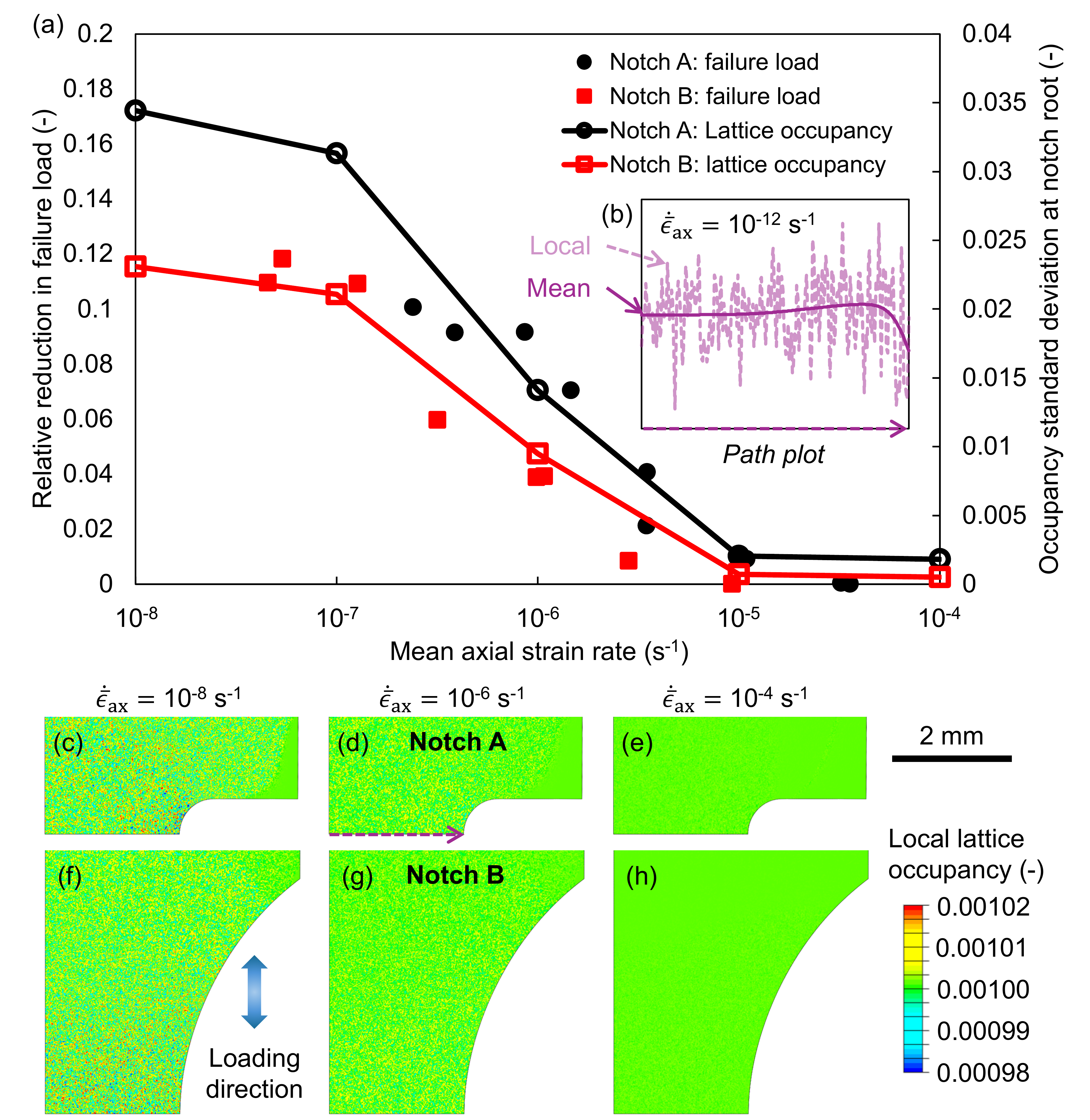}
  \caption{The linkage between hydrogen embrittlement, microstructure-driven redistribution of hydrogen, and strain-rate in 316L. (a) Correlation between experimental data \cite{TORIBIO1991} showing an increased reduction in failure load in notched samples (type A and B illustrated below) at lower strain-rates, and an increased level of microstructural redistribution of hydrogen over the same strain-rate range. (b) The relative distributions of mean (mm-scale) and local ($\upmu$m-scale) hydrogen redistribution in 316L (type A notch) at an extremely low strain-rate. Subfigures (c) - (e) and (f) - (h) show the distributions of local lattice occupancy in type A and B notches at 1\% equivalent strain for strain-rates 10$^{-8}$, 10$^{-6}$, and 10$^{-4}$ s$^{-1}$, respectively. The loading direction shown corresponds to subfigures (c) to (h).}
  \label{SRS_experiment_comparison}
\end{figure}

316L and other austenitic stainless steels are frequently reported as having low susceptibility to HE and associated microstructure sensitivity due to very low diffusivity and high hydrogen solubility, particularly when compared to ferritic steels \cite{MARTIN2020}. However, there are several examples of reduced yield strength, reduced ultimate tensile strength, and reduced failure load in hydrogen environments \cite{SAN2012}, particularly at low strain-rates. Despite this, comprehensive datasets are scarce. To establish a connection between multiscale hydrogen diffusion and HE, Toribio's failure load reduction data \cite{TORIBIO1991} is utilised for comparison in Figure \ref{SRS_experiment_comparison}, as it covers a wide strain-rate range. 
Toribio \cite{TORIBIO1991} considered similar notch geometries to those studied in this paper, referred to as notch A (where the notch radius is 0.5 mm) and notch B (where the notch radius is 5 mm) henceforth. Interestingly, the data shows a notable difference between notches in terms of the strain-rate range over which the greatest changes occur (adopting the mean axial strain definition). In Figure \ref{SRS_experiment_comparison} (a), this experimental data is compared semi-quantitatively with the predicted local lattice occupancy standard deviation at each notch root in the strain-rate range $10^{-8}$ s$^{-1}$ to $10^{-4}$ s$^{-1}$. Relative reduction in failure load is defined as $1-\sigma_{\mathrm{F}}^{\mathrm{H}}/\sigma_{\mathrm{F}}^0$, where $\sigma_{\mathrm{F}}^{\mathrm{H}}$ is the failure load in the presence of hydrogen, and $\sigma_{\mathrm{F}}^0$ is the failure load in air at the equivalent strain-rate. The predicted normalised deviation of lattice occupancy, $\frac{\Delta\theta_{\mathrm{L}}}{\bar{\theta}_{\mathrm{L}}}$, is plotted on a separate vertical axis in Figure \ref{SRS_experiment_comparison} (a), and demonstrates remarkable agreement with failure load reduction in terms of its strain-rate dependent evolution, and in capturing the comparably subtle differences between notches A and B. Toribio \cite{TORIBIO1991} claimed that at lower strain-rates than considered in their study, the failure load reduction will saturate. Whilst there is not enough data to substantiate this claim, $\frac{\Delta\theta_{\mathrm{L}}}{\bar{\theta}_{\mathrm{L}}}$ is shown to approach two distinct saturation levels at very low strain-rates in notches A and B (saturation is achieved where $\dot{\epsilon}_y\leq10^{-9}$ s$^{-1}$). At the high strain-rate end, there is further overlap between experiment and model, in that effectively no failure load reduction is observed where the local hydrogen redistribution is minimised. Over the strain-rate range $10^{-8}$ s$^{-1}$ to $10^{-4}$ s$^{-1}$, there is negligible millimetre scale (notch) diffusion for either case owing to the very low diffusivity of 316L (see Table \ref{MultiProps}). It is not until strain-rates reach as low as $10^{-12}$ s$^{-1}$ that millimetre scale (mean) diffusion levels become comparable to saturated microstructure redistribution levels, as shown in Figure \ref{SRS_experiment_comparison} (b). Hence, the millimetre scale hydrostatic stress distribution differences between notches A and B (see Figure \ref{EPinterface}) should have no influence on HE in such cases, whilst von Mises stress differences yield differences in local redistribution, which is implicated in microscale diffusion-driven HE. These results strongly indicate that elastic-plastic anisotropy driven diffusion at the microstructural level plays a major role in HE of austenitic steels and possibly other metallic materials. For further illustration, Figure \ref{SRS_experiment_comparison} (c) - (e) and (f) - (h) show local lattice occupancy distributions across the entire strain-rate range for notches A and B, respectively. These subfigures highlight and compare the very visible local redistribution of hydrogen against the virtually nonexistent millimetre scale diffusion (which is not visible at even the lowest strain-rate shown here due to the selected colourbar range). 

\begin{table}[htb]
\centering
\caption{Multiscale mechanical and hydrogen diffusion properties for 316L and nickel.}
\begin{tabular}{ l l l l }
\hline 
 Property & 316L & Nickel & Units\\
 \hline
 \multicolumn{3}{l}{Mechanical response \cite{KANG2010,BARZDAJN2018,ARYA2021}}\\
 \hline
 $E$                                & $2.00\times10^{11}$     & $2.05\times10^{11}$     & Pa           \\
 $\sigma_{\mathrm{y}}^0$            & $2.20\times10^{8}$      & $3.00\times10^{7}$      & Pa           \\
 $\sigma_{\mathrm{y}}^{\mathrm{p}}$ & $2.52\times10^{8}$      & $6.06\times10^{7}$      & Pa           \\
 $M_{\mathrm{T}}$                   & $3.07$                  & $3.07$                  & -            \\
 $G^{i\in[1,M]}$                    & $1.26\times10^{11}$     & $1.27\times10^{11}$     & Pa           \\
 $b^{i\in[1,M]}$                    & $2.54\times10^{-10}$    & $2.49\times10^{-10}$    & m            \\
 $\rho_{\mathrm{SSD},0}$            & $2.45\times10^{10}$     & $1.00\times10^{11}$     & m$^{-2}$     \\
 $k$                                & $7.00\times10^{13}$     & $1.20\times10^{14}$     & m$^{-2}$     \\
 \hline
 \multicolumn{3}{l}{Micromechanical response contributors \cite{LEDBETTER1984,HACHET2018}}             \\
 \hline
 $G$                                & $7.00\times10^{10}$     & $7.60\times10^{10}$     & Pa           \\
 $K$                                & $1.03\times10^{11}$     & $1.56\times10^{11}$     & Pa           \\
 $C_{11}$                           & $2.05\times10^{11}$     & $2.66\times10^{11}$     & Pa           \\
 $C_{12}$                           & $1.38\times10^{11}$     & $1.50\times10^{11}$     & Pa           \\
 $C_{44}$                           & $1.26\times10^{11}$     & $1.27\times10^{11}$     & Pa           \\
 \hline
 \multicolumn{3}{l}{Hydrogen transport \cite{ELMUKASHFI2020,BRASS2006,HILL1955,ATRENS1977}}            \\
 \hline
 $D_{\mathrm{L}}$                   & $1.704\times10^{-16}$   & $1.720\times10^{-13}$   & m$^2$s$^{-1}$\\
 $W_{\mathrm{dis}}$                 & $-2.2431\times10^{-20}$ & $-2.9890\times10^{-20}$ & J            \\
 $V_{\mathrm{L}}$                   & $1.1591\times10^{-29}$  & $1.0950\times10^{-29}$  & m$^{3}$      \\
 $d_{\mathrm{g}}$                   & $1.00\times10^{-5}$     & $1.00\times10^{-5}$     & m            \\
 \hline
\end{tabular}
\label{MultiProps}
\end{table}

\begin{figure}[t!]
  \centering
  \includegraphics[width=1\linewidth]{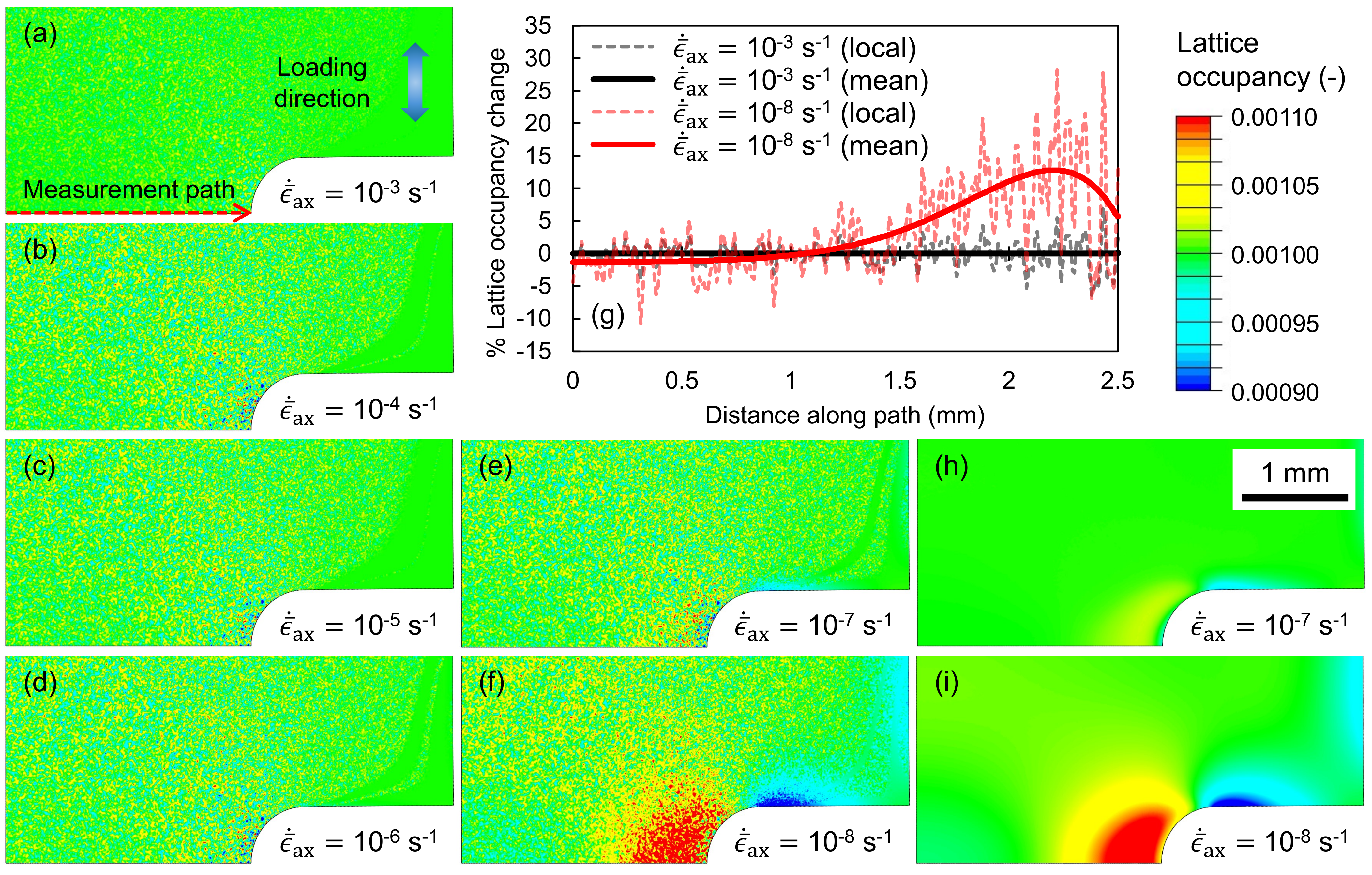}
  \caption{(a) - (f) Distributions of local lattice occupancy at 1\% equivalent strain in notched pure nickel test specimen (type A) across a wide range of strain-rates. (g) Comparison of local and mean lattice occupancy distributions at extreme ends of considered strain-rate range. Mean lattice occupancy distributions are shown in (h) and (i) for strain-rates of 10$^{-7}$ and 10$^{-8}$ s$^{-1}$, respectively.}
  \label{Ni_SRS}
\end{figure}

By contrast, Figure \ref{Ni_SRS} presents various distributions of local and mean lattice occupancy terms at 1\% mean axial strain across a strain-rate range ($10^{-8}$ s$^{-1}$ to $10^{-3}$ s$^{-1}$) which spans the onset of microstructural redistribution to the dominance of millimetre scale (mean) diffusion in nickel. Subfigures \ref{Ni_SRS} (a) to (f) present local lattice occupancy distributions across notch A in order of increasing redistribution and reducing strain-rate, showing saturation of microstructural redistribution around $\dot{\epsilon}_{\mathrm{y}}=10^{-6}$ in (d), and subsequently the increasing influence of millimetre scale diffusion on the total local content in (e) and (f). Subfigures \ref{Ni_SRS} (h) and (i) show the corresponding mean lattice occupancy fields after loading at the two lowest strain-rates. 

Figure \ref{Ni_SRS} (g) shows the distribution of local and mean lattice occupancies along the measurement path shown in Figure \ref{Ni_SRS} (a) at the highest and lowest strain-rates considered. At the high end, the local lattice occupancy change is shown to fluctuate by up to 7\%. At the low strain-rate however, microstructure-driven fluctuations reach up to 15\% near the notch root, which is combined with mean redistribution of the same amount, yielding up to a 30\% increase in local hydrogen concentration.

\section{Discussion}

Using coupled CPFE and hydrogen transport models, we capture the relative effects of elastic and plastic anisotropy on the steady state redistribution of hydrogen at microstructural length scales. Under fully elastic and elastic-plastic loading conditions and in the long-term limit, diffusible hydrogen concentration is shown to follow the Gaussian distribution, as expected \cite{HUSSEIN2021}, but also to broaden by up to 50\% relative to the elastic case with the introduction of mean plastic strains on the order of 1 - 2\%. Analytical sub-modelling reveals that the diffusible hydrogen distribution  is directly linked to the hydrostatic stress, which also follows a Gaussian distribution. An interesting question concerns the effect of length scale on mechanical response. Our present model could be extended to account for this via GND hardening, which may yield non-Gaussian hydrostatic stress and hydrogen concentration distributions that are likely to be more difficult to capture analytically. Moreover, in the small strain regime considered here, hydrogen redistribution is assumed to be dominated by diffusion, i.e., so that the contribution of trapped hydrogen is small by comparison. For high strains ($>$ 10\%), it has been reported that the total hydrogen distribution becomes log-normal \cite{HUSSEIN2021} due to significant trapping in regions with high levels of plasticity. This effect could be readily incorporated to the current framework by linking trap evolution to GND and SSD densities. 

In the present study, the goal is to lay the foundations for microstructure-informed analytical modelling of HE. We have presented a framework which extracts and generalises the key relationships for FCC metals without convoluting various contributing factors. Hence, there is still much scope for the development of micromechanical hydrogen transport models, which may rely upon lower length scale modelling methods \cite{ZHOU2019} for parameterisation, particularly in the absence of robust experimental data. We also hope that this research will spur on the experimental HE community to develop experiments to obtain the required data for validation and calibration of any such extensions to the current modelling framework. 

In the development and application of analytical sub-models for microscale hydrostatic stress distributions and the transient localisation of hydrogen in this paper, there are several prerequisites. The first is that the microstructure should be untextured and equiaxed as this forms the basis of Fokin and Shermergor's internal stress model \cite{FOKIN1968}. Moreover, for multimodal grain size distributions, the characteristic length scale approximation in Equations (\ref{deltathetaL}) and (\ref{l_c}) would require some extension using e.g., a Fourier transform based approach for weighted superposition of hydrogen redistribution at multiple length scales. Secondly, for fully elastic loading, the models are only valid for polycrystals with cubic symmetry, since the mean hydrostatic stress (which influences stress distributions in non-cubic polycrystals \cite{ELSHAWISH2024}) is currently unaccounted for. Hence, there is also the opportunity to extend the current framework to model materials with other crystal structures, e.g., for zirconium alloys, in which the localisation of hydrogen is particularly important for hydride formation \cite{MOTTA2019}. Importantly, the current framework demonstrates that with relevant properties, the key characteristics of microstructure-driven hydrogen redistribution may be captured at millimetre length scales without using computationally intensive CPFE models (except for validation). Figure \ref{SRS_experiment_comparison} demonstrates the importance of this, as the localisation of hydrogen at the microstructural level in 316L is strongly implicated in HE of test specimens with millimetre scale geometries across a broad range of strain-rates (see Figure \ref{SRS_experiment_comparison}). Whilst Toribio \cite{TORIBIO1991} reported that the failure load differences between sharp and blunt notch types were very small, this subtlety correlates exceptionally well with the predicted transient microscale localisation of hydrogen. In the absence of hydrogen, differences in failure mode between sharp and blunt "Bridgman" notches \cite{BRIDGMAN1952} have been widely reported \cite{MIRZA1996,INOUE2021}. Specifically, sharp notch geometries generate an increase in hydrostatic stress relative to the deviatoric stress, resulting in brittle fracture behaviour where ductile fracture is otherwise observed \cite{LEI2015}. For 316L however, whose tensile properties are weakly affected by hydrogen, ductile fracture was reported in air and hydrogen environments for both notch geometries. It is interesting to note that despite the reported increase in notch strengthening with decreasing notch radius in 316L (due to local plasticity and stress relief) \cite{WILLIAMS1963}, in the presence of hydrogen, the lower radius notch geometry was shown to promote a more dramatic (relative) load reduction \cite{TORIBIO1991}. This indicates that HELP is unlikely to have contributed to the reported failure load reduction data studied here, since greater load reductions are associated with an increased hydrostatic stress, as compared to the deviatoric stress. 

For materials with very low hydrogen diffusivity, our results suggest that microstructure is the dominant factor affecting hydrogen redistribution (at least for monotonic loading at realistic engineering strain-rates). The reverse of this argument may suggest that BCC metals, which have higher diffusivity values \cite{SMIRNOVA2023} are less sensitive to microstructure in terms of HE. However, owing to their comparably low hydrogen solubility, any localisation due to microstructure is also likely to have an important effect. Pure nickel has a hydrogen diffusivity that is nearly three orders of magnitude greater than that of 316L. Hence, in Figure \ref{Ni_SRS}, which presents the combined microscale and macro-scale hydrogen redistribution fields in nickel, it is shown to be sensitive to diffusion at both length scales within the strain-rate range considered. Whilst the spatial gradients of hydrostatic stress generated by microstructure are higher than those created by notches due to lower length scale, the greatest overall hydrostatic stress differences are at the millimetre scale (see Figure \ref{EPinterface} (c)). Hence, since the diffusivity for nickel is higher than for 316L, millimetre scale diffusion effects are more pronounced. These findings suggest that for materials with very high diffusivity values (e.g., $>$ 10$^{-6}$ m$^2$s$^{-1}$), millimetre scale diffusion may dominate under multiaxial loading conditions. In the current study of nickel, however, the combined effects of microstructure and notch geometry are shown to drive significant hydrogen redistribution with near equal respective contributions. The ability to account for these combined effects is a key advantage of the current multiscale modelling approach (as opposed to considering local and mean hydrogen redistribution in isolation).

\begin{figure}[htb]
  \centering
  \includegraphics[width=0.55\linewidth]{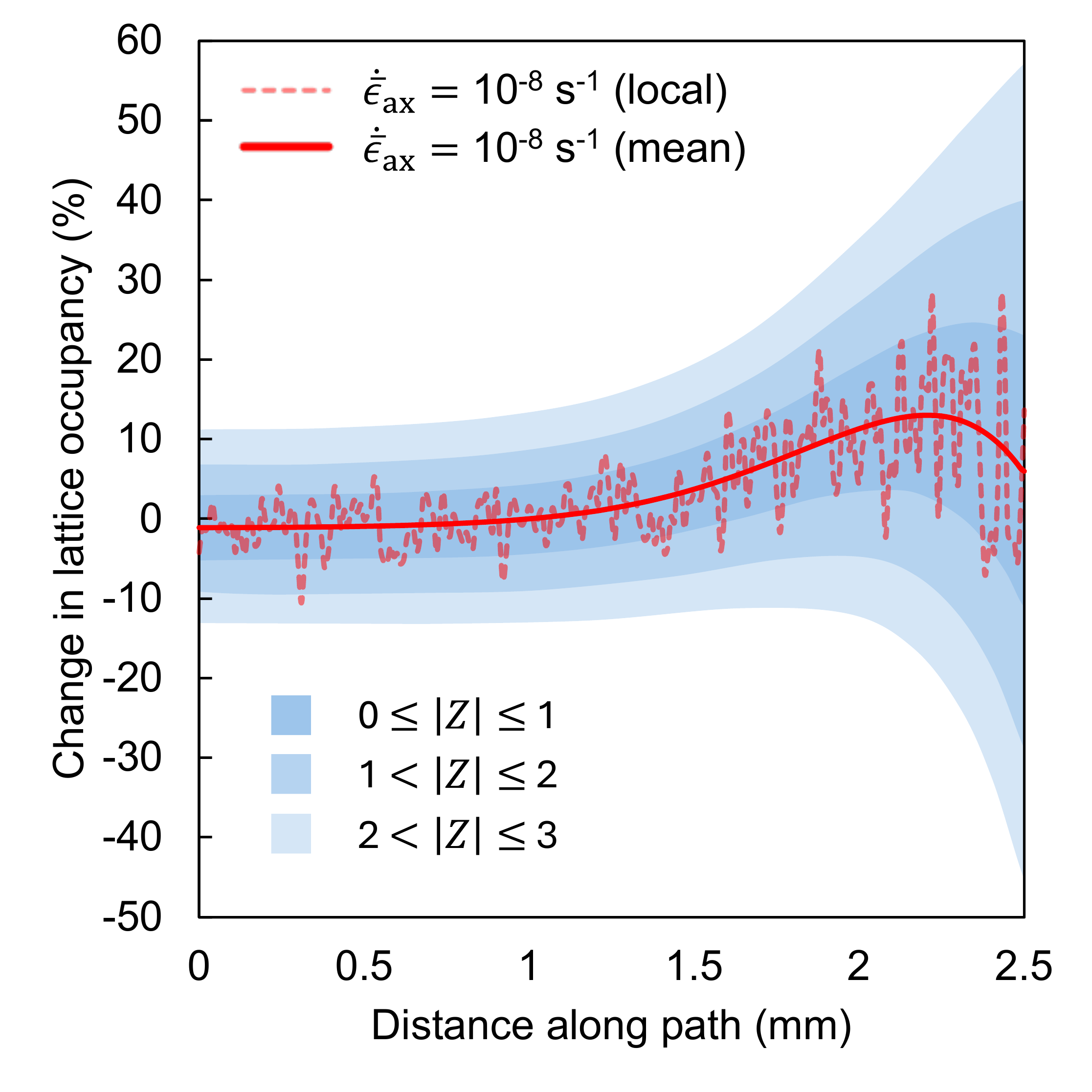}
  \caption{Comparison of local and mean lattice occupancy distributions at low strain-rate in nickel (see Figure \ref{Ni_SRS} (g)) with overlay of corresponding local occupancy bounds based on various $Z$ ranges.}
  \label{Statistics_illustration}
\end{figure}

An alternative to the present stochastic modelling approach is to carry forward the full statistical information about the microscale hydrogen distribution within each macroscopic element, rather than attempting to reproduce a representative hydrogen distribution. Hence, rather than attempting to reproduce a representative distribution of hydrogen, it may be more conservative to select a fixed value for $Z$ throughout the model so that the greatest anticipated local hydrogen concentration can be estimated within each continuum element. A reproduction of the low strain-rate data from Figure \ref{Ni_SRS} (g) is presented in Figure \ref{Statistics_illustration} with an overlay of various $Z$ range contours. It illustrates that for the path considered, most values lie within 1 standard deviation of the mean, with some between 1 and 2 standard deviations, and virtually none between 2 and 3 standard deviations. Hence, it would be most conservative, i.e., a worst case scenario to set $Z$ = 3 everywhere throughout the model, while eliminating any potential mesh sensitivity concerns. On the other hand, having a random (and representative) distribution of hydrogen may facilitate the prediction of HE-driven crack nucleation or HEDE using e.g., cohesive zones modelling \cite{ELMUKASHFI2020,JEMBLIE2017}. HEDE is often seemingly random and abundant (from a continuum perspective) \cite{WADA2023}, and otherwise impossible to capture under uniaxial loading conditions.

Figure \ref{Statistics_illustration} highlights another important aspect of the current multiscale modelling approach which is that the maximum local hydrogen content simultaneously depends upon mean hydrostatic and mean von Mises stress distributions. Whilst plastic relaxation near the notch root yields a local reduction in mean hydrostatic stress, the von Mises stress continues to increase. Hence, from a solely continuum modelling perspective, notch root plasticity will yield an apparent reduction in local hydrogen content, whereas microstructure-driven redistribution can produce a direct countereffect, particularly for $Z>2$. 

In broader terms, multiscale models such as this may have important implications for material or microstructure design. For example, the key microstructure design principles learned from the analytical models are that to minimise hydrogen localisation under loading, isotropic elasticity and resistance to plastic deformation are favourable.

\section{Conclusions}

In this paper we have developed a stochastic multiscale finite element modelling framework that captures the combined effects of stress state, microstructure heterogeneity, strain-rate, and diffusivity on the overall redistribution of hydrogen for continuum level analyses. The framework was evaluated using case studies in 316L stainless steel and pure nickel under multiaxial loading conditions, the results of which brought forth the following conclusions:
\begin{itemize}
  \item For elastically anisotropic cubic polycrystals, only the deviatoric stress (represented here using mean von Mises stress) affects the microscale redistribution of hydrogen due to the orientation invariance of hydrostatic stress for crystals with cubic symmetry. 
  \item In both materials, plasticity is shown to increase local (microstructural) gradients of hydrostatic stress, and hence to accelerate the grain scale redistribution of hydrogen. 
  \item For materials with very low diffusivity, e.g., 316L, microstructure-driven hydrogen redistribution will dominate over continuum level effects for industrially-relevant strain-rates. The strain-rate dependence of microscale redistribution is also shown to correlate strongly with the strain-rate dependence of HE in 316L, suggesting it is a key contributor to HE.  
  \item In nickel, an interplay between microstructural and millimetre scale hydrogen redistribution is predicted over the strain-rate range $\dot{\bar{\epsilon}}=10^{-8}$ to $10^{-3}$ s$^{-1}$. A short (grain scale) diffusion path leads to microstructure dominance at high strain-rates, whilst the large hydrostatic stress variation manifested by a continuum scale notch feature is shown to yield levels of localisation on par with microstructure at low strain-rates. 
\end{itemize}

\section*{Acknowledgements}

The authors would like to acknowledge Rolls-Royce plc. for their financial and technical support in this project (grant number RR/UTC/89/9 BPC 189). We would particularly like to thank Chris Argyrakis, Louise Gale, and Duncan Maclachlan for their input. 

\bibliographystyle{unsrt}
\bibliography{References}

\end{document}